\documentclass{tlp}

\usepackage[latin1]{inputenc}

\usepackage{amsmath}  %%por \align
\usepackage{theorem}  %% por \theoremstyle
\usepackage{verbatim}
\usepackage{url}

\usepackage{txfonts}
\DeclareFontFamily{OT1}{txss}{}
\DeclareFontShape{OT1}{txss}{m}{n}{ <-> s * [0.9] txss }{}
\DeclareFontFamily{OT1}{txtt}{}
\DeclareFontShape{OT1}{txtt}{m}{n}{ <-> s * [0.9] txtt }{}

\usepackage{pstricks,pst-node,pst-tree,pst-3d}

\usepackage{latexsym}
\usepackage{stmaryrd} %% Por el rayo!
\newcommand{\abort}{\lightning}

\newcommand{\todo}[1]{\begin{center}\textbf{\fbox{\Large TODO}:} #1\end{center}}
\renewcommand{\todo}[1]{}

\newcommand{\bibdir}{../../../babylon}

\newcommand{\smalltt}[1]{\texttt{#1}}%OJO

\newcommand{\bart}{\bar{t}}
\newcommand{\barx}{\bar{x}}

\newcommand{\quitar}[1]{}

\newcommand{\HP}{\mathsf{H}}

\newcommand{\Rules}{\mathit{Rules}}

\newcommand{\VS}{\mathit{VS}}

\newcommand{\FS}{\mathit{FS}}

\newcommand{\Env}{\mathit{Env}}

\newcommand{\EE}{\textrm{\bf E}}

\newcommand{\FF}{\textrm{\bf F}}
\newcommand{\RR}{\textrm{\bf R}}
\newcommand{\semb}[1]{[\![{#1}]\!]}

\newcommand{\seq}{\doteq}

\newcommand{\valiun}[1]{\mathbf{E}[\![{#1}]\!] I}

\newcommand{\need}{\ensuremath{\Leftarrow}}

\newcommand{\menor}{\ensuremath{\sqsubseteq}}
\newcommand{\mayor}{\ensuremath{\sqsupseteq}}

{
\theoremstyle{plain}
\theorembodyfont{\rmfamily}
\newtheorem{example}{Example}%[chapter]
\newtheorem{definition}[example]{Definition}
\newtheorem{lemma}[example]{Lemma}

\newtheorem{theorem}[example]{Theorem}

\newtheorem{myrule}{Rule}
\newtheorem{myproof}{Proof of Rule}

}
\newcommand{\ra}{\rightarrow~}
\newcommand{\fl}{\ensuremath{\rightarrow}}

\newcommand{\bydef}{\stackrel{\rm{\tiny def}}{=}}

\newcommand{\syntb}[1]{[\![{#1}]\!]}

\newcommand{\fname}[1]{\mathit{#1}}

\newcommand{\Bool}{\fname{Bool}}
\newcommand{\True}{\fname{True}}
\newcommand{\False}{\fname{False}}
\newcommand{\Two}{\fname{Two}}
\newcommand{\PP}{\fname{PP}}
\newcommand{\Nat}{\fname{Nat}}
\newcommand{\length}{\fname{length}}
\newcommand{\plus}{\fname{plus}}
\newcommand{\foldl}{\fname{foldl}}
\newcommand{\Succ}{\fname{Succ}}
\newcommand{\Zero}{\fname{Zero}}
\newcommand{\false}{\fname{false}}

\newcommand{\ppname}[1]{\fname{#1}}
\newcommand{\ppis}[1]{\ppname{is}\fname{#1}}
\newcommand{\ppglb}{\ensuremath{\sqcap}}
\newcommand{\pplub}{\ensuremath{\sqcup}}
\newcommand{\pphnf}[1]{{\ppname{hnf}\fname{#1}}}
\newcommand{\ppnf}[1]{{\ppname{nf}\fname{#1}}}
\newcommand{\ppprj}[2]{{\ppname{prj}#1\fname{#2}}}
\newcommand{\ppany}{\ppname{any}}
\newcommand{\ppnothing}{\ppname{nothing}}
\newcommand{\pphnflist}{\ppname{hnfList}}
\newcommand{\ppspine}{\ppname{spine}}
\newcommand{\ppnflistnat}{\ppnf{ListNat}}
\newcommand{\pphnfnat}{\pphnf{Nat}}
\newcommand{\ppnfnat}{\ppnf{Nat}}
\newcommand{\ppnil}{\ppname{nil}}
\newcommand{\ppcons}{\ppname{cons}}
\newcommand{\ppzero}{\ppname{zero}}
\newcommand{\ppsucc}{\ppname{succ}}
\newcommand{\samelength}{\ppname{samelength}}
\newcommand{\hnf}{\fname{hnf}}
\newcommand{\spine}{\fname{spine}}

\newcommand{\ppty}[1]{#1\rightarrow\Two}

\title{Demand Analysis with Partial Predicates%
\footnote{This is the extended version of a paper
          accepted for publication in a forthcoming special
          issue of \emph{Theory and Practice of Logic Programming}
          on Multiparadigm and Constraint Programming
          (Falaschi and Maher, eds.)
          Appendices are missing in the printed version.}}
\shorttitle{Demand Analysis with Partial Predicates}

\author[Mariño, Herranz, Moreno-Navarro]
       {Julio Mariño, Ángel Herranz and 
        Juan José Moreno-Navarro\\
        {~~~~~~~~~~Universidad~Politécnica~de~Madrid~~~~~~~~~}}

\submitted{4 January 2004}
\revised{1 March  2005}
\accepted{13 January 2006}

\begin{document}
\maketitle

%% A B S T R A C T
\begin{abstract}
In order to alleviate the inefficiencies caused by the interaction of
the logic and functional sides, integrated languages may take
advantage of \emph{demand} information --- i.e.~knowing in advance
which computations are needed and, to which extent, in a particular
context. 
This work studies \emph{demand analysis} -- which is closely related to 
\emph{backwards strictness analysis} -- in a semantic framework of
\emph{partial predicates}, which in turn are constructive realizations
of ideals in a domain. 
This will allow us to give a concise, unified presentation of 
demand analysis,
to relate it to other analyses based on abstract 
interpretation or strictness logics, some hints for the implementation,
and, more important, to prove the 
soundness of our analysis based on \emph{demand equations}.
There are also some innovative results. 
One of them is that a set constraint-based analysis has been derived
in a stepwise manner using ideas taken from the area of program
transformation. 
The other one is the possibility
of using program transformation itself to perform the analysis, 
specially in those domains of properties where algorithms based on 
constraint solving are too weak.
%The other one is a proposal to lift partial predicates in order to
%cope with higher order demand properties. 
%The full paper will include code generation schemes that will show the
%applicability of the information obtained from the analysis.
%% ¿No debería ser automático?
\end{abstract}


%\label{demand}
% Aquí tiene que venir la nueva entradilla del capítulo
\section{Introduction}
Although the main idea of declarative programming is to use 
mathematical elements for programming, the area is split in two main 
paradigms based on the subset of mathematics they are focused on:  
functional programming (functions: lambda calculus) and 
logic programming (predicate  logic). However it is obvious 
that both paradigms have a common core and can be seen as  
different faces of a single idea.
  
\emph{Functional-logic languages} aim at
bringing together the advantages of functional programming and logic programming,
see \cite{hanusurvey,jjsurvey}, i.e.\ from functional programming
they take higher-order
features, polymorphic types, lazy evaluation, etc., while
logic programming provides partial information, 
constraints, logical variables, search, etc. The language Curry
\cite{curryreport:0.8}
is the de facto standard of functional-logic languages.

Probably, the combination of the last 
mentioned features of each paradigm (laziness, search) 
seems the more problematic to achieve in
an efficient implementation. In other words, 
for executing a program it may be necessary to evaluate a 
functional-like expression containing uninstantiated (i.e. 
existentially quantified) logic variables.

The operational principle proposed for this situation is narrowing. 
Roughly speaking, narrowing guesses an instantiation for these variables. 
Functional nesting, nondeterminism, semantic unification, 
functional inversion, and lazy evaluation, which are key part for the 
expressiveness of functional-logic programs, are supported by this 
operational mechanism.
In order to apply the general idea of narrowing to functional-logic 
languages, a functional-logic program is considered as a set of 
rewrite rules (plus some additional restrictions described later). 

However, narrowing by itself is not enough: 
a brute-force approach to finding all the solutions 
would attempt to unify each rule with each nonvariable subterm of 
the given equation in every narrowing step. Even for small programs
a huge search space would result.
Therefore, an additional aspect to take into account
in the implementation of functional-logic languages is the definition of an 
appropriate narrowing strategy.
This  strategy  should  be  sound  (i.e.,  only  correct  solutions  
are computed)  and  complete  (i.e.,  all  solutions  or  more  
general  representatives  of  all  solutions  are computed).  

Many narrowing strategies for limiting the size of the search 
space have been proposed, but we are interested on those with a 
lazy behaviour. To preserve completeness, see \cite{babeljfp},
a lazy narrowing step is applied at outermost positions with 
the exception that inner arguments of a function are evaluated, 
by narrowing them to their head normal forms, if their values 
are required for an outermost narrowing step.  
This property can only be ensured by looking-ahead on the 
rules tried in following steps.  Unfortunately, a potentially 
infinite number of substitutions could arise but,
in the case of inductively sequential programs it is possible to compute
the property in an efficient way.  This is the idea of needed narrowing 
introduced in \cite{antoy:echahed:hanus:2000:jacm}.  
The paper also proves completeness and  
optimality of the strategy.

But beyond this remarkable contribution to the implementation of 
functional-logic languages two problems remain: (i) the strategy is 
optimal with respect to the length of derivations but not in the 
size of the search space and additional improvements could be 
achieved, and (ii) it is defined only for a restricted class of 
programs (inductively sequential).

\subsection{Demand analysis}
Our proposal is to use a static analysis to improve the look-ahead 
approach. The analysis is able to extract demand information from a
functional-logic program.  This information can be used
to guide and improve needed narrowing (thus cuting the search space
and avoiding reevaluations).  Additionally, we have some other advantages:
transform nonsequential programs into sequential ones, following the
ideas of \cite{padl00} where strictness analysis is used;
safe replacement of strict equality by
unification; implementation of default rules \cite{lazydef}; 
improvement of the accuracy of groundness
analysis; translation into Prolog \cite{JMM92,wflp98sloth}, etc.

\emph{Demand analysis} was introduced in \cite{JMM92} and then used in
\cite{MKM93,HM93,MHM93} as a way to improve the compilation of
functional-logic programs. The essential idea was to perform backwards
strictness analysis.  The proposed solution consisted in
generating, for a given program, a set of so called \emph{demand
  equations} that were solved in a domain of regular trees
(\emph{demand patterns}).
A strong point of demand patterns compared to existing strictness analyzers
based on abstract interpretation was that they allowed, at least 
theoretically, inference in an infinite domain of properties.

The aim of this paper is to provide a semantic framework
(\emph{partial predicates}) for demand analysis, to prove the
correctness of demand equations and to introduce a novel approach to
the implementation of analyzers. In fact, the whole process is
simplified and optimized: while it is easier to reason about demand
analysis with partial predicates, the implementation is also simpler
because we can reuse a lot of work already done in partial evaluation
tools. Moreover, we claim that the method can be applied to different
contexts with a similar success.

Partial predicates can be used to specify demand analysis as 
well as other classical analyses. 
One of the important point of this formalism is that it can be
expressed by functional-logic programs. This has important
consequences, allowing for reasoning
about demand properties (for instance checking and inference of them) 
by using program transformation techniques.

%% \subsection{Applications of demand analysis}
%% Let us conclude the introduction by motivating possible applications of demand
%% analysis to the compilation of functional-logic programs. 
%% Notice that they go
Applications of demand analysis go
beyond the usual applications of strictness analysis in functional
programming, where strictness is used for trying to provide bigger 
computations that can be computed eagerly.
Advances on this subject can have effects in getting a more efficient 
low level implementation, and in making easier and profitable the 
parallelization of programs.
In functional-logic programming the gain in effiency means 
that some computations are not reevaluated
or even that a wider class of programs can be used.

%\item [The Reevaluation Problem] \index{reevaluation}
\paragraph{The Reevaluation Problem.} 
There is an efficiency problem caused by the interaction of
laziness and backtracking which are, in some sense, antagonistic in
nature. The former tries to delay some computations while the second
drives different computations through a tree of branching paths. The
problem appears clear -- if we place some computation beyond the
branching point there exists the risk that the evaluation may be
performed several times. This can be rather annoying, because laziness
is intended to save work, not to waste it.

\begin{figure}
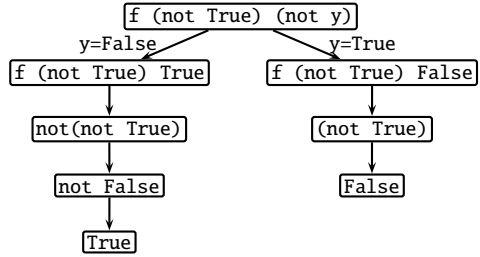

  \begin{minipage}{0.4\linewidth}
\begin{verbatim}
not False = True
not True  = False

f x False = not x
f x True  = x

?- f (not True) (not y)
\end{verbatim}
  \end{minipage}
  \begin{minipage}{0.45\linewidth}
\begin{small}
\begin{center}
\psset{arrows=->,framearc=.2}
\psset{levelsep=0.75cm}
\newcommand{\Treebox}[2][]{\Tr[#1]{\psframebox[framesep=1pt]{\smalltt{#2}}}}
\pstree[treemode=D]{\Treebox{f (not True) (not y)}}{
  \pstree{\Treebox{f (not True) True}\ncput{\smalltt{y=False~~~~~~~~~~}}}{
    \pstree{\Treebox{not(not True)}}{
      \pstree{\Treebox{not False}}{
        \Treebox{True}
      }
    }
  }
  \pstree{\Treebox{f (not True) False}\ncput{\smalltt{~~~~~~~~~~y=True}}}{
    \pstree{\Treebox{(not True)}}{
      \Treebox{False}
    }
  }
}
%\nput[nodesep=0.5]{0}{s2}{\smalltt{x }}
%\nput[nodesep=0.5]{0}{s3}{\smalltt{x }}
%\end{minipage}
\end{center}
\end{small}
\end{minipage}
\caption{Reevaluation example.}
\label{fig:reevaluation}
\end{figure}

The toy example in Figure~\ref{fig:reevaluation} shows how combining
lazy evaluation and backtracking can lead to the repeated evaluation
of delayed redexes.  Observe that the redex \texttt{(not True)} is
needed for the final result but, due to the outermost reduction
strategy, its evaluation is delayed and evaluated twice, in different
branches of the search tree.
In this simple example, this is not serious, but in general, the
redex reevaluated could have an expensive operation and the
reevaluation can take place not only two, but an unbounded (even
infinite) number of times.

\begin{figure}[t]
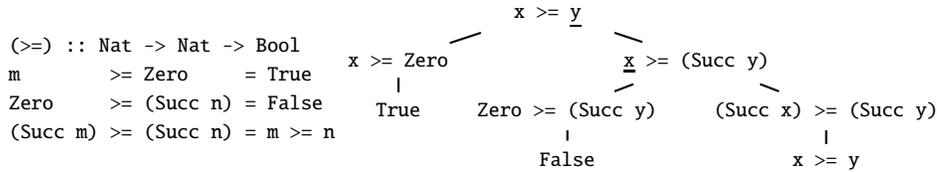

\begin{minipage}[c]{0.35\linewidth}
\begin{verbatim}
(>=) :: Nat -> Nat -> Bool
m        >= Zero     = True
Zero     >= (Succ n) = False
(Succ m) >= (Succ n) = m >= n
\end{verbatim}
\end{minipage}
\begin{minipage}[c]{0.64\linewidth}
\begin{tt}
\begin{small}
\pstree[nodesep=4pt,levelsep=5ex]{\TR{{~~x >= \underline{y}~~}}}{
  \pstree{\TR{x >= Zero}}{
    \TR{True}
  }
  \pstree{\TR{\underline{x} >= (Succ y)}}{
    \pstree{\TR{Zero >= (Succ y)}}{
      \TR{~~~~False~~~~}
    }
    \pstree{\TR{(Succ x) >= (Succ y)}}{
      \TR{~~~~x >= y~~~~}
    }
  }
}
\end{small}
\end{tt}
\end{minipage}
\caption{Sample definition and its associated definitional tree.}
\label{samples}
\end{figure}

\paragraph{Sequentiality Analysis.} 
One of the stages in compiling the code for a function definition in a
lazy language is to decide (i) which arguments need reduction to
perform the matching against the patterns in the left hand sides of
the rules, and (ii) in which order are these arguments to be
reduced. For instance, the code for the \textit{greater or equal}
predicate (Figure~\ref{samples}) needs to obtain a topmost data
constructor for the second argument in order to choose a rule.  If
this constructor is \texttt{Zero}, the only match is with the first
rule and no evaluation is needed on the first argument, but if it is
\texttt{Succ}, the first argument needs to be examined in order to
choose between the second and third rules.

Sloth\footnote{\texttt{http://babel.ls.fi.upm.es/software}\,.}
\cite{wflp98sloth}, our implementation of Curry, uses
\emph{definitional trees}~\cite{antoy:echahed:hanus:2000:jacm}  
as the intermediate structure
to store these decisions. Figure~\ref{samples} shows the definition
for \texttt{(>=)} and the definitional tree obtained from
it. Underlined positions are those where the branching of the
decisions are done. Next section will provide a definition of this
concept but the graphical notation could be enough at this point. 
Observe
that arguments are not necessarily examined in a left to right order.
%
%% Ejemplo del ``mayor o igual''. Incluye código y deftree. 

Sometimes, it is difficult -- or impossible -- to discover
a definitional tree from the left hand sides alone.
A typical example is the merging of two sorted lists:
%
%\begin{small}
\begin{verbatim}
merge :: [Nat] -> [Nat] -> [Nat]
merge [] ys = ys         merge (x:xs) (y:ys) | x <= y = x:(merge xs (y:ys))
merge xs [] = xs                             | x >= y = y:(merge (x:xs) ys)
\end{verbatim}
If we just look at the left hand sides there is no way of building a
decision tree like the one for \texttt{(>=)}, but looking at the
right hand sides we immediately see that both arguments to
\textit{merge} are demanded. A programmer would write, in fact, a
modified version where the second rule is rewritten as
\begin{verbatim}
merge (x:xs) [] = x:xs
\end{verbatim}
instead. In our paper~\cite{padl00} we show that for the vast majority
of programs, the information obtained from type inference and
demand analysis can help a compiler to generate sequential
definitional trees for programs even when they are not syntactically
sequential.

\paragraph{Avoiding Redundant Tests in Target Code.} 
A substantial part of the overhead in implementing a lazy
functional-logic language is precisely due to the code that implements
the lazy evaluation of arguments (this was, in fact, what motivated
research on strictness analysis of functional programs in the first
place).  This is particularly evident when looking at the translation
scheme into Prolog of Sloth \cite{wflp98sloth}, but the problem
appears also in abstract machine based implementations.

Demand information can help to generate, for most functions, eager
code, with a drastic effect on efficiency. In the case of the
translator to Prolog, this allows an almost verbatim translation with
very little overhead.

\paragraph{Relation with Freeness and Sharing Analysis.}
\emph{Freeness}\index{freeness analysis} and
\emph{sharing}\index{sharing analysis}%
~\cite{%marriot93precise,
jacobs:langen:1992:jlp,muthukumar91combined}
are two operational properties of logic programs
which have been extensively studied. Freeness analysis can tell
whether a free variable present in a goal does not get bound during its
evaluation. Sharing analysis can tell whether two variables present
in a goal will not eventually share some structure.
Freeness and sharing are important, for instance, for the parallel
execution of logic programs. Moreover, freeness can be useful to
detect deterministic computations in functional-logic programs.
%(see \ref{determinacy}). % OJO!!

Freeness and sharing are much more difficult to study in a lazy
functional-logic language than in Prolog. From an operational point
of view, the techniques used for Prolog can predict whether a given
expression may bind some variables \emph{provided that} this
expression gets \emph{actually evaluated}. 

From a denotational point of view, this connection can be explained
because the lazy semantics can be seen as a restricted form of
free-variable semantics where all variables are collapsed into one
symbol ($\bot$). In this setting, variable propagation (freeness) and
$\bot$-propagation (strictness) share a common mechanism
\cite{mitesis}.

%% The semantics needed for this is developed in chapter
%% \ref{semantics} \departamento{ and the relation is studied in chapter \ref{sharfree}}.

%% \firstreport{
%% \item [Application to Default Rules] 
%% \departamento{
%% \emph{Juanjo, ¿podrías detallar esto?}
%% }
%% }
%%   \index{default rules}

\paragraph{Transformation of Strict Equality into Unification.}
The behaviour of the equality operator is slightly different in logic programming
and in functional-logic programming. While in the first case the expression $x = t$, where $x$
is a free variable, can be dealt with by assigning $t$ to
$x$ (provided that $x$ does not occur in $t$), in functional-logic programming it is
necessary to ensure that $t$ can be evaluated to a \emph{total} value,
i.e.\ a term with no undefined subterms, otherwise the whole
expression will be undefined.  When logic programs are translated into
functional-logic programs, this has two undesired effects: the expression cannot be resolved
in constant time -- which precludes the use of Prolog techniques such
as difference lists -- and the computed answers
can be unnecessarily detailed.

The connection with strictness is twofold. On one hand, totality is a
much stronger condition than not being undefined, which makes rules
with equality expressions a source of useful information for a
strictness analyzer.
On the other hand, the same techniques employed to avoid redundant
tests when applying strictness analysis can be used to avoid the
test that $t$ does not contain an occurrence of a defined function
symbol, which is a sufficient condition to perform the assignment.

%\end{description}

%%%%%%%%%%%%%%%%%%

%%%%%%%%%%%%%%%%%%%%%%%%%%%%%%%%%%%%%%%%%%%%%%%%%%%%%%%%%%
% Aquí se acaban los problemas, así que dejamos un espacio
% de párrafo libre.
\paragraph{}
Some of these problems can be tackled by nonstandard implementation
architectures like \emph{memoization} or 
\emph{bottom-up execution}~\cite{magic95} for the reevaluation
problem, or \emph{parallel definitional trees}~\cite{AEH97,Genius96}
to cope with the lack of sequentiality but, in practice, these methods
introduce their own overheads and actual implementations are based
either on extensions of abstract machines for the execution of
functional languages or logic languages, or on the translation to another
declarative language, like Prolog, see \cite{JMM92,wflp98sloth}. 
This is why we chose to
attack the problem at compile time or, in other words, how our
research on demand analysis began.

\subsection{Paper Organization}
Section~\ref{preliminaries} introduces the subset of the Curry language
we are going to use along the paper as well as the operational 
semantics used, namely narrowing.
Section~\ref{ppredicates} introduces the formalism of partial
predicates, their structure and how demand properties can be
represented by using them.  Section~\ref{checking} discusses the
problem of checking and inferring demand properties of a program. 
Different
sublattices of demand properties, with increasing complexity, are
introduced. Checking is demonstrated in an abstract way, by means of
syntactic transformations (fold/unfold).  Later, the harder problem of
inferring demand properties is considered, firstly 
(Section~\ref{sec:inference}), in an abstract way and then 
(Section~\ref{dac}) in connection with a particular analysis tool.
Correctness of the aforementioned demand equations is shown there as
well as the algorithms to compute approximate solutions to them.  Code
generation based on the information from the analyzer is discussed in
Section~\ref{codegen} and a few experimental results showing the
feasibility of the method are shown in Section~\ref{results}. Some
related work is discussed in Section~\ref{related} and open issues
in Section~\ref{future}.  Finally, Section~\ref{demand:conclusion} concludes.
In order to make this paper as self-contained as possible,
\ref{app:semantics} includes the reference denotational
semantics for the kernel language.
\ref{app:proof} contains some proofs that have
been removed from the printed version due to lack of space.

\section{Preliminaries}\label{preliminaries}

This section is devoted to fix the subset of Curry that will be used
along the paper. The operational semantics assumed is discussed too.

\subsection{Kernel Language}
The language of choice is largely immaterial but a little syntax is
needed in order to keep some coherency throughout the paper. We will
use a simplified version of the functional-logic language Curry
\cite{curryreport:0.8}, basically a language of recursion equations,
with a Haskell-like syntax.  In the sequel we assume some knowledge of
functional-logic languages and Curry operational semantics.

We assume a ranked set $\mathit{TC} = \bigcup_n \mathit{TC}^n$ of
\emph{type constructors} $K$ and a countably infinite set $\mathit{TV}$ of
\emph{type variables} $\alpha$.
Any data type is uniquely denoted by an algebraic term
$\tau \in \mathcal{T}(\mathit{TC}\cup\mathit{TV})$
or a function type $(\tau_1 \rightarrow \tau_2)$.
Next, we assume a set $\mathit{DC} = \bigcup_n \mathit{DC}^n$ of typed
\emph{data constructors} $C$, a countably infinite set $\mathit{VS}$ of
\emph{variable symbols} $x$, and a set $\mathit{FS}$ of \emph{function
symbols} $f$ with declared principal type
$f: \tau_1 \rightarrow \dots \rightarrow \tau_n \rightarrow \tau$
where $\tau$ is not a function type.  $TC$, $DC$, $VS$ and $FS$ are disjoint.
The \emph{arity} of a data constructor $C \in \mathit{DC}^n$ is $n$ and
is denoted $\mathit{ar}(C)$.
In practice, type and data constructors are both defined via
\textit{data} declarations of the form
\begin{displaymath}
  \texttt{data}~K~\alpha_1 \dots \alpha_l =
  C_1~\tau_{11}~\dots~\tau_{1m_1}|~ \dots ~|~ C_n~\tau_{n1}~\dots~\tau_{nm_n} \,.
\end{displaymath}
A type constructor $\Bool$ with
data constructors \textit{True} and \textit{False} is always assumed.
Expressions are given by the grammar
\begin{displaymath}
  e ::= C ~|~ x ~|~ f ~|~ e_1~e_2 \,.
\end{displaymath}
Expressions must be well typed.
%% in the sense of a Milner system.
A program is a set of defining rules of the form
\begin{displaymath}
  f~e_1 \dots e_n~\mathtt{=}~[b \rightarrow] e \,.
\end{displaymath}
%% Here $n$ is the (rule) arity of $f$, denoted $\mathit{ar}(f)$.
The optional condition $b$ of type $\Bool$ is called the {\em guard} of
the rule.  Several restrictions are imposed on the rules in a program
in order to ensure confluence of reduction, and the following are used
somewhere in the paper: (i) for every rule $(l~\mathtt{=}~r)$, $l$ is
a \emph{pattern}, i.e.\ it has a single function symbol at its top and
no variable occurs twice in $l$; (ii) rules must be well typed; (iii)
for every pair of program rules $(l_1 = r_1)$, $(l_2 = r_2)$, if $l_1$
and $l_2$ have a unifier $\sigma$ then $\sigma(r_1) = \sigma(r_2)$;
(iv) free variables -- i.e.~those occurring in the right hand side but
not in the left hand side -- are allowed only if their rightmost
occurrence is in the guard. Moreover, they must be of first order type.
Observe that we are not forbidding overlaps. The set of rules defining
function symbol $f$ in program $P$ is denoted $\Rules_P(f)$.
%% nor free variables.

When looking at the syntactic shape of the left hand sides of defining
rules, a total application $(f~e_1~\dots~e_n)$ is treated as the
algebraic term $f(e_1,\dots,e_n)$ and then the standard notation for
positions and substitutions is used. A position is a
string of natural numbers that identifies a path to a subterm in a
term. The expression $t|_p$ denotes the subterm of $t$ at position
$p$, i.e.\ $f(e_1,\dots,e_n)|_{i.p} = e_i|_p$ and $t|_\epsilon = t$, with
$\epsilon$ the empty string.  Replacement of $t|_p$ by $t'$ is
abbreviated $t[t']_p$.
The topmost symbol of term $t$ is denoted $root(t)$.

A denotational semantics for the kernel language can be found in
\ref{app:semantics}.  
Although not strictly necessary to understand the
techniques proposed here, this is the ultimate foundation for the
validity of the equations and inequalities used and supports the
validity of the fold/unfold transformations.

\subsection{Narrowing}
The fundamental computation mechanism of functional-logic languages is 
narrowing. 
Informally, to \emph{narrow} an expression $e$ means 
to apply a substitution that makes it reducible, and then reduce it. 
An expression 
$e$ narrows to $e'$ with substitution $\sigma$, if $p$ is 
a nonvariable position of $e$, 
$l = r$ is a variant of a program rule sharing no variables with $e$, 
and $\sigma$ is a substitution such that $\sigma(l) = \sigma(e|_p)$, 
and $e' = \sigma(e[r]_p)$.

As we have mentioned, unrestricted application of the narrowing rule
is too nondeterministic and many strategies have been proposed to
improve this. From an expressiveness point of view, we prefer those
with a lazy behaviour because functions can be defined more
independently, without interaction among them, thus increasing
modularity and reusability and allowing programming techniques like
infinite objects. A general description of lazy narrowing for
functional-logic languages can be found in \cite{babeljfp}.

The task of a narrowing strategy is the computation of the step, 
or steps, that must be applied to a term. 
A narrowing strategy suitable for functional-logic languages 
must be sound, complete, and efficient. 
The intuition behind the soundness and the completeness of a strategy 
when  the initial term of a derivation is an equation containing unknown 
variables is easy to describe: 
Soundness guarantees that any instantiation of the 
variables computed by the strategy is a solution 
of the equation. Completeness ensures that for any solution of the equation, 
the strategy computes another solution which is at least as general.

However, efficiency is more difficult to state. 
As usual, the goal is to minimize the 
overall time and memory consumed when finding one 
or all the values of an expression. 
In the narrowing context, it is related with the length 
of the derivations, and, specially, with the size 
of the search space.
Basically, the two factors affecting the efficiency of a strategy are: 
(i) unnecessary steps should be avoided, and 
(ii) steps should be computed without consuming unnecessary resources. 
Lazy narrowing  steps try to be applied at outermost positions but 
to preserve completeness (see \cite{babeljfp}) inner arguments 
of a function are evaluated, 
by narrowing them to their head normal forms, if their values 
are required for an outermost narrowing step.

In general, a strategy cannot easily determine if a computation is 
unnecessary without look-ahead. 

The strategy used in Curry is \emph{needed narrowing}%
~\cite{antoy:echahed:hanus:2000:jacm}, 
a lazy strategy where the 
program is translated into a set of definitional trees, one for every 
function symbol being defined. Definitional trees are given by the grammar
\begin{displaymath}
DT~~::=~\mbox{branch (\emph{Pattern}, Pos  [, DT ]}^+) 
        ~|~ \mbox{rule \emph{Rule}} ~|~ \mbox{or }(DT [, DT]^+) \,,
\end{displaymath}

\noindent
where  \emph{Pattern}  stands for patterns made up 
of data constructors and different 
variables as in the left hand sides of program rules,  
\emph{Pos}  are positions defined    
in  the  standard way and  \emph{Rules}  are program rules. 
Trees without  \emph{or} nodes   
are called (inductively) sequential, otherwise 
they are parallel  definitional trees. 

Given an expression $e$ and a set of definitional trees for the 
defined symbols of the program, a position in $e$ can be chosen 
to apply narrowing. This is done by first looking for 
an outermost application $f e_1 \dots e_n$ where  $f$  is a 
defined function symbol, and then descending some  $e_i$  according 
to  $f$'s definitional tree. 
Therefore, needed narrowing with sequential definitional trees 
establishes an efficient algorithm for implementing the look-ahead
for required evaluations. In \cite{antoy:echahed:hanus:2000:jacm} 
the formal definition is
given but also an interesting property is shown: the strategy is 
optimal with respect to the length of derivations.

To overcome the restriction to inductively sequential programs, 
other strategies have been proposed. 
For instance, the strategy of \cite{fraguasloogenartalejo93} 
is based on some form of generalized 
definitional trees. The completeness of this strategy is unknown. 
These strategies are \emph{demand driven},
which informally means the following: a subterm $v$ of a term 
$t$ is evaluated if there is a rule $R$ potentially applicable 
to $t$ that demands the evaluation of $v$.

The lack of well-defined strategies with provable properties motivated 
alternative efforts for computations in this class. In our case,
we use the information provided by demand analysis to guide the computation.
The analysis, its formal properties and the use of demand information
for implementing efficiently functional-logic languages are the goal
of the following sections.

%%%%%%%%%%%%%%%%%%%%%%%%%%%%%%%%%%%%%%%%%%%%%%%%%%%%%%%%%%%%%%%%%%%%%%%%%%%
\section{Partial Predicates}\label{ppredicates}
This section is devoted to introduce the formalism of \emph{partial
  predicates}.  Informally speaking, they are logic predicates that
represent the degree of evaluation of an expression (demand
properties).  The main feature is that they can be described using the
language under analysis, so the language itself is used to abstract some
properties of a given program.  This fact is essential for their use
as an analysis tool, as will be shown later.%
\footnote{From this point on, notation based on \emph{domain theory}
  is extensively used, and a denotational semantics for the kernel
  language is assumed. Readers less familiar with these topics are
  referred to~\cite{handbooktcs-b}, Chapters 11 and 12.}

\begin{definition}[Partial Predicate]
  Let $\Two$ be a two point domain; a \emph{partial predicate} $\pi$
  defined on type $\tau$ is any continuous map %
  $\pi \in \tau \rightarrow \Two$ which can be defined in the kernel
  language.  In the following, the type scheme $\PP~\alpha$ will be
  used as synonymous with $\alpha~\fl~\Two$.
\end{definition}

A two point domain $\Two$ is isomorphic to the subset of the domain
$\Bool = \{\True, \False, \bot\}$ after removing
$\False$\footnote{Hence the name of partial predicates.} and can be
defined in the kernel language:
\begin{verbatim}
data Two = True
\end{verbatim}
The definition of conjunction and disjunction functions in $\Bool$ can
be restricted to this domain:
\begin{verbatim}
(&&), (||) :: Two -> Two -> Two
True && True = True
True || y = True;                   x || True = True
\end{verbatim}
Observe that disjunction in $\Two$ is given by a parallel definition. 
While this could be problematic in case of trying to execute the program, 
it is not the case in our context as long as the definitions of 
the predicate transformers will mainly be used for program transformation.

Every partial predicate $\pi$ of type $\PP~\tau$ represents subsets of
the domain $\tau$:
\begin{displaymath}
  \pi^{-1}(\True) = \{ x \in \tau ~|~ \pi(x) = \True\} \,.
\end{displaymath}

\begin{example}[Peano naturals]
  Some of the examples throughout the paper will make use of a data
  type for Peano naturals:
\begin{verbatim}
data Nat = Zero | Succ Nat
\end{verbatim}
  A pair of predicates $\pphnfnat$ and $\ppnfnat$ can be introduced:
\begin{verbatim}
hnfNat, nfNat :: PP Nat
hnfNat Zero = True;                 hnfNat (Succ _) = True
nfNat Zero = True;                  nfNat (Succ n) = nfNat n
\end{verbatim}
  The former yields $\True$ when its argument is evaluated enough to
  identify its topmost constructor.  The latter yields $\True$ when
  its argument is evaluated to normal form.  Observe that %
  $\pphnfnat~n = \bot \Leftrightarrow n = \bot$.
\end{example}

\begin{example}[Partial predicates $\pphnflist$ and $\ppspine$]
  \label{ex:pphnflist}
  Less trivial are those predicates that can be used to express
  properties of polymorphic types.  Consider, for instance,
  $\pphnflist$ or $\ppspine$ in the domain of polymorphic lists:
\begin{verbatim}
hnfList, spine :: PP [a]
hnfList [] = True;                  hnfList (h:ts) = True
spine [] = True;                    spine (h:ts) = spine ts
\end{verbatim}
  Analogously to the example above, the former yields $\True$ when its
  list argument is evaluated enough to identify its topmost
  constructor.  The latter yields $\True$ when the argument is
  evaluated enough to reach the end of the list.  In both cases, the
  degree of definition of the elements in the list is immaterial.
  Observe again that $\pphnflist~xs = \bot \Leftrightarrow xs = \bot$.
\end{example}

\begin{example}[Partial predicates $\ppany$ and $\ppnothing$]
  \label{ex:ppany}
  A pair of polymorphic partial predicates $\ppany$ and $\ppnothing$
  can be defined for all types:
\begin{verbatim}
any, nothing :: PP a
any x = True;                       nothing x = nothing x
\end{verbatim}
\end{example}

\subsection{Demand Typings}
\label{sec:demandtypings}
Partial predicates can be used to express a great number of program
properties, including classic strictness. For instance, suppose we are
interested in proving $f \in [\tau_1] \fl\ \Nat$ strict:\footnote{A
  formal definition of \emph{strict} can be found in
  Definition~\ref{def:strict}.}
\renewcommand{\iff}{\Leftrightarrow}
\begin{displaymath}
  \begin{array}{lll}
    f~\textrm{is strict} & \iff~ f~\bot = \bot ~\iff~ \forall x.~ x = \bot \Rightarrow f~x = \bot\\
    & \iff~ \forall x.~ f~x \sqsupset \bot \Rightarrow x \sqsupset \bot\\
    & \iff~ \forall x.~ \pphnfnat (f~x) = \True \Rightarrow \pphnflist~x = \True\\
    & \iff~ \forall x.~ \pphnfnat (f~x)\sqsubseteq \pphnflist~x & \iff~ \pphnfnat \circ f \sqsubseteq \pphnflist
  \end{array} \,.
\end{displaymath}
% Observe that the replacement of implication ($\Rightarrow$) by
% ordering ($\sqsubseteq$) is correct: $\bot \sqsubset \True$.
So the property `$\!f$ is strict' is equivalent to $\pphnfnat \circ f
\sqsubseteq \pphnflist$.

If we are interested in studying how much information is needed in a
function's argument in order to obtain a certain amount in the result,
this can be generalized to properties of the form %
$\pi_2 \circ f \sqsubseteq \pi_1$.

\begin{definition}[Demand Properties, Demand Types and Demand Typings]
  \label{def:demandproperty}
  A \emph{demand property} is an inequality of the form
%%  \begin{eqnarray*}
%%    \pi_2 \circ f & \sqsubseteq & \pi_1
%%  \end{eqnarray*}
  $\pi_2 \circ f \sqsubseteq \pi_1$
  where $\pi_1$ and $\pi_2$ are partial predicates defined on the
  domain and codomain types of $f$, respectively.  It is denoted
  in the following way:
  \begin{eqnarray*}
    f:\pi_1 \need \pi_2 & \bydef & \pi_2 \circ f \sqsubseteq \pi_1 \,.
  \end{eqnarray*}
  $\pi_1 \need \pi_2$ is a \emph{demand type} and %
  $f : \pi_1 \need \pi_2$ is a \emph{demand typing}.
\end{definition}
Demand typings $f : \pi_1 \need \pi_2$ are usually read `$\!f$ demands
$\pi_1$ to its argument in order to give a result as evaluated as
$\pi_2$.'  Our previous strictness example would be rewritten %
$f : \pphnflist \need \pphnfnat\;$.
It is rather simple to show that for any partial predicate
$\pi^{-1}(\True)$ is an ideal -- it is the inverse image of a closed
set.  In fact, `partial predicate' can be taken as synonymous with
`computable ideal.'  The lattice of partial predicates induces
another on demand types $\pi_1 \need \pi_2\,$: covariantly on $\pi_1$
and contravariantly on $\pi_2$.

\subsection{Polymorphism}
One advantage of partial predicates over other approaches to program
analysis, such as abstract interpretation on finite domains, is a
natural treatment of polymorphism.  Being written in source code they
have the same type constraints and expressiveness.

Some partial predicates presented so far are polymorphic and the same
can be said of the predicate transformers to be introduced below.

\begin{definition}[Predicate Transformers]
  \emph{Predicate transformers} are higher order functions that take
  partial predicates as (part of) its argument and yield partial
  predicates as result.
\end{definition}

Data constructors can be seen as predicate transformers, in
particular, polymorphic constructors can be represented by polymorphic
transformers.

\begin{definition}[Constructor Predicate Transformer]
  \label{def:ppconstructors}
  Consider a type definition of the form
  \begin{displaymath}
    \texttt{data}~K~\alpha_1~\dots~\alpha_l ~=~ \dots~|~C_i~\tau_1~\dots~\tau_{m_i}~|~\dots
  \end{displaymath}
  The \emph{constructor predicate transformer} $c_i$\footnote{We are
    not actualy overloading data constructor names but just changing
    first letter of the name to lowercase.} associated with each data
  constructor is defined as follows:
  \begin{tabbing}
    \=\kill
    \>\texttt{$c_i$ :: $\PP$ $\tau_1$ $\fl\dots\fl$ $\PP$ $\tau_{m_i}$ $\fl$ $\PP$ ($K~\alpha_1~\dots~\alpha_l$)}\\
    \>\texttt{$c_i$ p$_1$ \dots\ p$_{m_i}$ ($C_i$ x$_1$ \dots\ x$_{m_i}$)}
    = \texttt{(p$_1$ x$_1$) \&\& \dots\ \&\& (p$_{m_i}$ x$_{m_i}$)} \,.
  \end{tabbing}
\end{definition}

Interesting partial predicates associated to data constructors and
data types can be defined by using constructor predicate transformers
(e.g. Definition~\ref{def:ppis} and Definition~\ref{def:pphnf}).

\begin{definition}[Matching Predicates]
  \label{def:ppis}
  For every data constructor $C$ the partial predicate $\ppis{C}$
  defined
  \begin{tabbing}
    \=\kill
    \>\texttt{$\ppis{C}$ :: $\PP$ ($K~\alpha_1~\dots~\alpha_l$)}\\
    \>\texttt{$\ppis{C}$ = ($c$ any $\dots$ any)}
  \end{tabbing}
  is called the \emph{matching predicate for constructor $C$}.
\end{definition}

\begin{definition}[Meets and Joins]
  The greatest lower bound operator $\ppglb$ (\verb./\.) and the least
  upper bound operator $\pplub$ (\verb.\/.) can be defined:
\begin{verbatim}
(/\), (\/) :: PP a -> PP a -> PP a
(p /\ q) x = (p x) && (q x);        (p \/ q) x = (p x) || (q x)
\end{verbatim}
\end{definition}

\begin{definition}[$\pphnf{}$ Predicates]
  \label{def:pphnf}
  For every type constructor $K$ with data constructors $C_1$, \dots,
  $C_n$, the partial predicate $\pphnf{K}$ defined
  \begin{tabbing}
    \=\kill
    \>\texttt{$\pphnf{K}$ :: $\PP$ ($K$ $\alpha_1$ $\dots$ $\alpha_l$)}\\
    \>\texttt{$\pphnf{K}$ = $\ppis{C_1}$ $\pplub$ $\dots$ $\pplub$ $\ppis{C_n}$}
  \end{tabbing}
  is the \emph{$\pphnf{}$ predicate of $K$}.
\end{definition}

\begin{example}
  \label{ex:pplist}
  Constructor predicate transformers and matching predicates
  associated with the list constructors \texttt{[] :: [$\alpha$]} and
  \texttt{(:) :: $\alpha$ $\fl$ [$\alpha$] $\fl$
    [$\alpha$]}\footnote{We are using $\fname{List}$, $\fname{Nil}$
    and $\fname{Cons}$ as names for \texttt{([])}, \texttt{[]} and
    \texttt{(:)}, respectively.} are:
\begin{verbatim}
nil :: PP a                                       isNil :: PP [a]
nil [] = True                                     isNil = nil
cons :: PP a -> PP [a] -> PP [a]                  isCons :: PP [a]
cons p q (x : xs) = (p x) && (q xs)               isCons = cons any any
\end{verbatim}
  The definition of $\pphnflist$ in Example~\ref{ex:pphnflist} can be
  rewriten
\begin{verbatim}
hnfList :: PP [a]                   hnfList = isNil \/ isCons
\end{verbatim}
\end{example}

\begin{example}
  \label{ex:pptup2}
  The constructor predicate transformer and the matching predicate
  associated with the tuple constructor \texttt{(,) :: $\alpha$ $\fl$
    $\beta$ $\fl$ ($\alpha$,$\beta$)}\footnote{We are using
    $\fname{Tup2}$ as the name for \texttt{(,)}.} are:
\begin{verbatim}
tup2 :: PP a -> PP b -> PP (a,b)                  isTup2 :: PP (a,b)
tup2 p q (x,y) = (p x) && (q y)                   isTup2 = tup2 any any
\end{verbatim}
  And the definition of $\pphnf{Tup2}$ is:
\begin{verbatim}
hnfTup2 :: PP (a,b)                 hnfTup2 = isTup2
\end{verbatim}
\end{example}

\begin{definition}[Cartesian Products]
  The \emph{cartesian product} of two partial predicates is defined as
  the constructor predicate transformer $\fname{tup2}$.  In the
  following we will use the infix operator \texttt{($\times$) ::
    $\alpha$ $\fl$ $\beta$ $\fl$ ($\alpha$,$\beta$)} as synonymous
  with $\fname{tup2}$.  The definition is generalised to arbitrary
  length tuples:
  \begin{tabbing}
    \texttt{(p$_1$ $\times$ $\dots$ $\times$ p$_n$) (x$_1$, $\dots$, x$_n$) = (p$_1$ x$_1$) \&\& $\dots$ \&\& (p$_n$ x$_n$)}
  \end{tabbing}
\end{definition}

\begin{example}[Projections on Tuples]
  A cartesian product implies the existence of projections.  We will
  show that there are actually two predicate transformers
  $\ppprj{1}{}$ and $\ppprj{2}{}$ with types
\begin{verbatim}
prj1 :: PP (a,b) -> PP a            prj2 :: PP (a,b) -> PP b
\end{verbatim}
  although their definition is somewhat special.  Mathematically, the
  following must hold:
  \begin{eqnarray*}
    (\ppprj{1}{}~p)^{-1} & = & \{ x ~|~ \exists y.~p(x, y) = \True\}\\
    (\ppprj{2}{}~p)^{-1} & = & \{ y ~|~ \exists x.~p(x, y) = \True\} \,.
  \end{eqnarray*}

  As has been said in the introduction, the kernel language does not
  forbid free variables in the equations. In fact, the denotational
  semantics of rules treats them via a least upper bound quantified
  over all the possible values in their type.  This means that an
  implementation of projections will be:
\begin{verbatim}
prj1 p x = p (x, y) -> True;        prj2 p y = p (x, y) -> True
\end{verbatim}
  This can be surprising to the reader more biased towards functional
  programming but is by no means strange if we look at Prolog or
  functional-logic languages as Curry itself.  In the special case that
  the variable being quantified is of a first order type, implementing
  such an equation is not a problem.
\end{example}

The extension of projections to arbitrary data types and data
constructors is trivial.  In particular projections on arbitrary
length tuples will be used in the following section.

\begin{definition}[Projections]
  \label{def:ppprojections}
  Given the data declaration scheme
  \begin{displaymath}
    \texttt{data}~K~\alpha_1~\dots~\alpha_l ~=~ \dots~|~C_i~\tau_1~\dots~\tau_{m_i}~|~\dots
  \end{displaymath}
  for each data constructor $C_i$ and for each $k \in
  \{1,\dots,m_i\}$, the \emph{projection partial predicate}
  $\ppprj{k}{C_i}$ is defined as\footnote{Observe that $k$ actualy
    expands: $\ppprj{1}{C_i}$, $\ppprj{2}{C_i}$, \dots}
  \begin{tabbing}
    \=\kill
    \>\texttt{$\ppprj{k}{C_i}$ :: $\PP$ $\tau_k$ $\fl$ $\PP$ ($K$ $\alpha_1$ 
      $\dots$ $\alpha_l$)}\\
    \>\texttt{$\ppprj{k}{C_i}$ p x = 
      p ($C_i$ x$_1$ $\dots$ x $\dots$ x$_{m_i}$) $\fl$ True} \,.
  \end{tabbing}
\end{definition}

\section{Checking and Inference of Demand Properties}
\label{checking}
This section studies some analyses and their domains of properties
under the prism of partial predicates.  The novelty of our approach is
that, by expressing partial predicates in a subset of the programming
language under analysis, a \emph{program transformation approach} is
feasible. All the examples below use the well known fold/unfold
transformations and are, thus, trivially correct for a language with a
lazy declarative semantics.

\subsection{Checking}
\label{subsec:checking}
The problem of deciding if a given partial predicate fulfills the
demand information of a given function is the \emph{checking problem}:
to prove that given $f$, $\pi_1$ and $\pi_2$, $f: \pi_1 \Leftarrow
\pi_2$ holds.

Concrete analyses fix a specific domain of checking properties, i.e.\
only a limited number of partial predicates are allowed.  
%
%% We can
%% define a methodology to solve the checking problem in one of these
%% concrete domains:
%% \begin{enumerate}
%% \item Associate a partial predicate to each element of the domain.
%% \item Define the checking property for the concrete domain.
%% \item Use equational reasoning (i.e.\ partial evaluation techniques)
%%   to prove particular checking properties in concrete applications.
%% \end{enumerate}
%
%% \todo{(AHN) Los revisores preguntan por la relación entre partial
%%   evaluation y equational reasoning.}
%
%% \begin{theorem}\emph{(Soundness of checking)}
%%   For any analysis described with the previous method, if a checking
%%   property is established, then the property holds in the denotational
%%   semantics of the language.
%% \end{theorem}
%% \begin{proof}
%%   Trivial from the soundness of equational reasoning.
%% \end{proof}
%
Let us show the translation of several domains of properties into the
language of partial predicates and exemplify checking by means of
equational reasoning.

\paragraph{Classic Strictness Analysis.}
The first attempt to mechanize strictness analysis is found in
\cite{Mycroft80}.  The aim is to detect when an argument can be safely
reduced in advance without affecting the termination properties of the
program.
\begin{definition}
  \label{def:strict}
  A function $f$ is said to be \emph{strict} iff $f~\bot = \bot$.
\end{definition}
Due to evident practical reasons, this definition is relaxed to cope
with the usual case of the argument belonging to a product type:
\begin{definition}
  A function $f$ is said to be \emph{strict in its $i$-th argument}
  iff
  \begin{displaymath}
    \forall~ x_1 \dots x_{i-1}~ x_{i+1} \dots x_n.~
    f(x_1,\dots,x_{i-1},\bot,x_{i+1},\dots,x_n) = \bot \,.
  \end{displaymath}
\end{definition}
As we have seen in Section~\ref{sec:demandtypings}, the first case can be expressed in our setting by
saying that the function demands an argument strictly more evaluated
than $\bot$ in order to produce a result strictly more evaluated than
$\bot$.  When both the argument and result types are constructed,
values other than $\bot$ can be finitely presented by enumeration of the
different constructors in the type:

\begin{lemma}
%   Let $K$ be a type constructor with data constructors
%   $C_1,\dots,C_n$,
  If $f$ is a function with type $f :: K\tau \fl
  K'\tau'$ then it is strict iff $f : \pphnf{K} \need \pphnf{K'} \,.$
\end{lemma}
\begin{proof}
  See proof in Section~\ref{sec:demandtypings} and replace
$\fname{Nat}$ and $\fname{List}$ with $K'$
  and $K$.
\end{proof}

Let us see an example that will also show how to use equational
reasoning in order to prove the demand typing correct.
\begin{example}
\label{ex:length}
Function \textit{length}
%\begin{small}
\begin{verbatim}
length :: [a] -> Nat
length [] = Zero;                     length (h:ts)  = Succ (length ts)
\end{verbatim}
%\end{small}
is strict.  This will be expressed as %
$\length : \pphnflist \need \pphnfnat$. Using the definitions of
$\pphnfnat$ and $\pphnflist$ seen before, we have to prove that
$\pphnfnat\circ\length\sqsubseteq\pphnflist$.  Then these equivalences
follow from the semantics of the kernel language:\footnote{Composition
  in Curry is denoted by with the symbol `\texttt{.}'\,.}
\begin{verbatim}
(hnfNat . length ) [] = hnfNat (length []) = hnfNat Zero = True
(hnfNat . length ) (x:xs) = hnfNat (length (x:xs))
                          = hnfNat (Succ (length xs)) = True
\end{verbatim}
Both \emph{rules} coincide with the equations of $\pphnflist$.
\end{example}

\begin{lemma}
  If $f$ is a function with type %
  \texttt{$f$ :: ($K_1\tau_1$,$\dots$,$K_n\tau_n$) $\fl$ $K\tau'$}
  then it is strict in the $i$-th argument iff %
  $f : \ppany~\times~\dots~\times~\pphnf{K_i}~\times~\dots~\times~\ppany \need \pphnf{K}\,$.
\end{lemma}
\begin{proof}
  \begin{displaymath}
    \begin{array}{ll}
      &
      \pphnf{K} \circ f \sqsubseteq \ppany~\times~\dots~\times~\pphnf{K_i}~\times~\dots~\times~\ppany\\
      \iff & 
      \forall~ x_1 \dots x_n.~
      \pphnf{K}(f(x_1,\dots,x_n))\sqsubseteq (\ppany~\times~\dots~\times~\pphnf{K_i}~\times~\dots~\times~\ppany)(x_1,\dots,x_n)\\
      \iff &
      \forall~ x_1 \dots x_n.~
      \pphnf{K}(f(x_1,\dots,x_n)) \sqsubseteq \pphnf{K_i}(x_i)\\
      \iff &
      \forall~ x_1 \dots x_{i-1}~ x_{i+1} \dots x_n.~
      \pphnf{K}(f(x_1,\dots,x_{i-1},\bot,x_{i+1},\dots,x_n)) \sqsubseteq \pphnf{K_i}(\bot)\\
      \iff &
      \forall~ x_1 \dots x_{i-1}~ x_{i+1} \dots x_n.~
      \pphnf{K}(f(x_1,\dots,x_{i-1},\bot,x_{i+1},\dots,x_n)) \sqsubseteq \bot\\
      \iff &
      \forall~ x_1 \dots x_{i-1}~ x_{i+1} \dots x_n.~
      \pphnf{K}(f(x_1,\dots,x_{i-1},\bot,x_{i+1},\dots,x_n)) = \bot\\
      \iff &
      \forall~ x_1 \dots x_{i-1}~ x_{i+1} \dots x_n.~
      f(x_1,\dots,x_{i-1},\bot,x_{i+1},\dots,x_n)) = \bot \,.
    \end{array}
  \end{displaymath}
\end{proof}

\begin{example}
  \label{ex:plus}
  Function \textit{plus}
\begin{verbatim}
plus :: (Nat, Nat) -> Nat
plus (Zero, m) = m;                   plus (Succ n, m) = Succ (plus (n, m))
\end{verbatim}
  is strict in its first argument: %
  $\plus : \pphnfnat \times \ppany \need \pphnfnat\,$. %
  Unfolding the definition of $\pphnfnat \times \ppany$ we get
  \begin{tabbing}
    \texttt{(hnfNat $\times$ any) (x,y) = (hnfNat x) \&\& (any y) = hnfNat x} 
  \end{tabbing}
  Unfolding $\pphnfnat \circ \plus$:
  \begin{tabbing}
    \texttt{(hnfNat . plus) (Zero,m) = hnfNat m $\sqsubseteq$ hnfNat Zero}\\
    \texttt{(hnfNat . plus) (Succ n,m) = True = hnfNat (Succ n)}
  \end{tabbing}
  so $\pphnfnat\circ\plus \sqsubseteq \pphnfnat \times \ppany$ and, by
  Definition~\ref{def:demandproperty}, %
  $\plus : \pphnfnat \times \ppany \need \pphnfnat\,$.
\end{example}

\paragraph{Wadler's Four Point Domain.}
Many interesting properties are related to the degree of evaluation
required on recursive data structures, like lists. For instance,
function \textit{length} needs a nil-ending list in order to produce a
definite result, but it is immaterial whether one or more of its
elements is undefined.

In \cite{Wadler87b} a four point abstract domain of degrees of
definiteness of monomorphic lists is introduced.  The domain, in
increasing order, can be given as:
\newcommand{\topin}{\ensuremath{\top\!\!\in\,}}
\newcommand{\botin}{\ensuremath{\bot\!\!\in\,}}
\begin{eqnarray*}
  \mathbf{4} &=& \{ \bot \sqsubset \infty \sqsubset \botin \sqsubset \topin \}
\end{eqnarray*}

\noindent
representing, respectively, the undefined list, any list with an
undefined suffix, finite lists with some undefined elements and total lists.
The original paper is not very formal and does not make clear that this
semantics for the four elements does not provide conjunctive nor disjunctive
closeness. This, of course, can be achieved if their semantics is changed
into:
\begin{displaymath}
  \begin{array}{rlrl}
    \topin & \textrm{any list} & \botin & \textrm{any nil ending list}\\
    \infty & \textrm{any list in head normal form} & \bot   & \textrm{an undefined list} \,.\\
  \end{array}
\end{displaymath}
These four levels of definiteness can be represented in our framework
by the three partial predicates: $\pphnf{List}$, $\ppspine$, and
$\ppnf{ListNat}$:
\begin{verbatim}
nfListNat :: PP [Nat]
nfListNat (x:xs) = (nfNat x) && (nfListNat xs);   nfListNat [] = True;
\end{verbatim}
The following can help demonstrate the transformational approach.
Let us prove $\length:\ppspine \need \ppnfnat$.  Let $R$ denote
$\ppnfnat\circ\length$.  Applying standard \emph{fusion} techniques
(see Section~\ref{sec:inference}) we successively obtain:
\begin{verbatim}
R []     = nfNat (length [])   = nfNat Zero = True
R (x:xs) = nfNat (length x:xs) = nfNat (Succ (length xs))
         = nfNat (length xs)   = R xs
\end{verbatim}
so we conclude that $R = \ppspine$.

\paragraph{Uniform Properties.}
\begin{figure}
\begin{pspicture}(0,0)(-6,6)
\psset{unit=1.7cm}
\psset{viewpoint=-1 -1 2}%iba mejor con el 1er diagrama
%\psset{viewpoint=-1 -1.5 2}
\ThreeDput(1,4,1){\rnode{A}{$\mathbf{1}$}}
\ThreeDput(1,3,1){\rnode{B}{$(c_3,\top)$}}
\ThreeDput(1,2,1){\rnode{C}{$(c_5,\top)$}}
\ThreeDput(1,1,1){\rnode{D}{$(c_4,\top)$}}
\ThreeDput(0,2,1){\rnode{E}{$(c_1,\top)$}}
\ThreeDput(0,1,1){\rnode{F}{$(c_2,\top)$}}
\ThreeDput(0,0,1){\rnode{G}{$(c_0,\top)$}}
\ThreeDput(1,3,0){\rnode{H}{$(c_3,\bot)$}}
\ThreeDput(1,2,0){\rnode{I}{$(c_5,\bot)$}}
%\ThreeDput(1,1,0){\rnode{J}{$(y,\bot)$}}      % = 0 !!
\ThreeDput(0,2,0){\rnode{K}{$(c_1,\bot)$}}
%\ThreeDput(0,1,0){\rnode{L}{$(\wedge,\bot)$}} % = 0 !!
\ThreeDput(0,0,0){\rnode{M}{$(c_0,\bot)$}}    % = 0
\psset{linewidth=0.2pt}
\ncline{A}{B}
\ncline{B}{C}
\ncline{C}{D}
\ncline{E}{F}
\ncline{F}{G}
\ncline{C}{E}
\ncline{D}{F}
\ncline{B}{H}
\ncline{C}{I}
%\ncline{D}{M}%{J} % redundante!
\ncline{E}{K}
%\ncline{F}{M}%{L} % redundante!
\ncline{G}{M}
\ncline{K}{M}%{L}
%\ncline{L}{M}
\ncline{H}{I}
%\ncline{I}{M}%{J} % redundante!
\ncline{K}{I}
%\ncline{L}{M}%{J}
\end{pspicture}
\caption{The lattice of uniform properties on lists.}
\label{reticulin}
\end{figure}
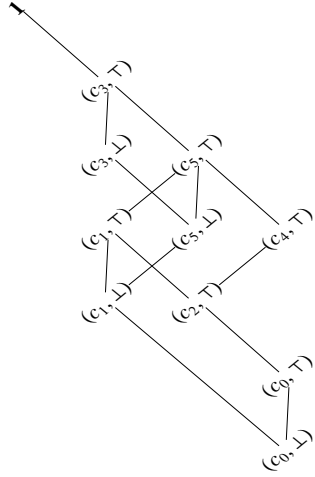
A number of proposals have been made to generalise Wadler's four point
domain to any algebraic datatype. Intuitively, the \emph{uniform
properties} of a data structure are those invariant under any (type
preserving) permutation of the elements of the structure. For
instance, if $p$ is an uniform property of lists of naturals, then
$ p (\Zero:(\Succ~\Zero):\bot) \Leftrightarrow p ((\Succ~\Zero):\Zero:\bot) $.

In \cite{Jensen94} a powerdomain construction for uniform properties
over algebraic datatypes is given, using modalities, along with a
strictness logic for reasoning about those properties. His domains are
able to express certain properties that do not appear in Wadler's,
like a list being empty or being finite with all their elements
undefined, etc. The formalism is rather involved and the strictness
logic does not lead very naturally to an implementation.

Trying to accommodate those ideas into our framework of partial
predicates we immediately see that the domains that arise in Jensen's
work correspond essentially to the \emph{folds} on a given datatype. A
fold on a data structure is a transformation that replaces every n-ary
constructor by an n-ary function.  Folds are generic programming
constructs in the sense that folds can be defined for every algebraic
datatype in an uniform way. For instance, folding lists is done using
the following higher order operator:
\begin{verbatim}
foldl :: (a -> b -> b) -> b -> [a] -> b
foldl f b [] = b;                   foldl f b (x:xs) = f x (foldl f b xs)
\end{verbatim}
Due to type restrictions, the number of partial predicates on natural
lists that can be defined as folds is limited. If only two degrees of
definiteness are considered for the naturals ($\bot$ and $\True$) that
leaves six possible combining functions of type $\Two\fl\Two\fl\Two\,$:
\begin{displaymath}
  \begin{array}{l@{~~~~}l@{~~~~}l}
    (c_0)~\lambda x.\ \lambda y.\ \bot          &
    (c_1)~\lambda x.\ \lambda y.\ x             &
    (c_2)~\lambda x.\ \lambda y.\ x \wedge y    \\
    (c_3)~\lambda x.\ \lambda y.\ \mathtt{True} &
    (c_4)~\lambda x.\ \lambda y.\ y             &
    (c_5)~\lambda x.\ \lambda y.\ x \vee y
  \end{array}
\end{displaymath}
times two values for the base case gives a lattice
(Figure~\ref{reticulin}) of at most 13 abstract values (adding
$\ppany$).\footnote{The notation $(c,v)$ stands for 
                   $\foldl~c~v\,$.
                   The exact cardinality of the abstract domain is 11,
                   as some of the combinations coincide: 
                   $\foldl~c_2~\bot = \foldl~c_4~\bot = \foldl~c_0~\bot\,$.}
A similar domain appears in~\cite{benton:phdthesis}.

The possibility of using catamorphisms (fold-like functions) on
algebraic types as a way of automatically constructing domains for the
analysis of programs has been studied and implemented
in~\cite{rey:2003:katana}.

\subsection{Inference}
\label{sec:inference}
Here we study the problem dual to checking, i.e.~how to
infer a partial predicate that describes (with reasonable accuracy)
demand information for a given function.
The concrete inference problem considered in this paper is the following:
given $f$ and $\pi_2$, to find the \emph{best} $\pi_1$ such that
$f: \pi_1 \Leftarrow \pi_2$ holds and, more important, to give a usable
representation.
The best $\pi_1$ such that $f: \pi_1 \Leftarrow \pi_2$ is
$ \pi_1 = \pi_2 \circ f $
and an explicit (recursive) definition can be obtained using the
transformational approach used for checking. However, giving a compact
representation of $\pi_1$ suitable for code generation can be
difficult and here is where a purely symbolic approach is better
suited than the program transformation one.
%
%% If computing $\pi_1$ is difficult, any $\pi'_1 \sqsupset \pi_1$ can be
%% considered a safe approximation.

%% This section considers the inference problem in a general setting,
%% while next one studies the correctness of a particular inference
%% method.
%% The inference process can be summarized in these three steps:

%% Es mejor comentar esto a posteriori, después de haber mostrado 
%% el ejemplo desarrollado, y ya enlazando con lo que viene después
%%
%% \begin{enumerate}
%% \item \textbf{Fusion}.  In this stage $\pi_1 = \pi_2 \circ f$ is
%%   partially evaluated in order to get a syntactic representation as
%%   simple as possible.

%%   \todo{(AHN) Los revisores preguntan por el significado de partial
%%     evaluation of $\pi_2 \circ f$.}

%% \item \textbf{Weakening} (optional).  If the recursive definition of
%%   $\pi_1$ obtained in step 1 is too complex, some of the terms in a
%%   right hand side of this definition can be replaced in such a way
%%   that this new predicate (let it be called $\pi_1'$) satisfies $
%%   \pi_1 \sqsubseteq \pi_1' \,. $
%% \item \textbf{Solving}.  Set constraint algorithms are applied in
%%   order to compute $(\pi_1')^{-1} (\mathtt{True})$.  These are more or
%%   less standard in the literature, so we will not discuss them here
%%   (see, for instance \cite{Aiken+93}).
%% \end{enumerate}

In order to get an informal understanding of the connection between
the program transformation and the symbolic approaches to inference,
let us revisit Example~\ref{ex:length} recast as an inference
problem:

\begin{example}

The original program is 
\begin{verbatim}
length []     = Zero;               length (h:ts) = Succ(length ts)
\end{verbatim}
We want to infer the degree of definiteness demanded $\pi_1$ on its
argument by a result in normal form, i.e. $\pi_1 = \ppnfnat\circ
\length$, so the following must hold:
\begin{eqnarray*}
    \pi_1~([])   &=& (\ppnfnat \circ \length)~([])\\
                 &=& \ppnfnat~(\Zero) ~=~ \True\\
    \pi_1~(h:ts) &=& (\ppnfnat \circ \length)~(h:ts)\\
                 &=& (\ppnfnat\circ\Succ)~(\length~(ts)) \,.
\end{eqnarray*}
It is easy to see that $\ppnfnat\circ\Succ$ simplifies to $\ppnfnat$ so
\begin{eqnarray*}
    \pi_1~(h:ts) &=& \ppnfnat~(\length~(ts))\\
                 &=& (\ppnfnat\circ\length)~(ts) ~=~ \pi_1~(ts) \,,
\end{eqnarray*}
resulting in a rather generative set of equations for $\pi_1$ (that
coincide with equations for $\ppspine$).
\end{example}

The following section shows a more systematic method to manipulate
partial predicates properties in a fully symbolic way. From the
program under analysis and the inference question, a set of
inequalities among symbolic representations of partial predicates is
generated, and is this set which is manipulated -- although the
meaning of these rewritings must mimic the original program transformations.

%% Ya han sido definidos
%The definition for the inverse constructors used in the example
%is:
%%
%\[
%  \begin{array}{l}
%   (\pi_1\texttt{:}\pi_2) (x\texttt{:}y) = \pi_1 x \wedge \pi_2 y \\
%   \texttt{[]} \texttt{[]} = \texttt{true}
%  \end{array}
%\]

%% %\noindent
%% To make a long story short, we have proved all the constraint solving
%% rules in the demand pattern approach to be equivalent to sound program
%% transformations involving partial predicates, i.e.\ demand pattern
%% analysis is correct.  In the next section, the process is developed
%% for our current tool, based on standard notions of set constraint 
%% analysis. 

%% Esto ya se explica más adelante
%%
%% \subsection{Weakening}
%% The generation of constraints in \cite{MKM93} and \cite{MHM93} is not
%% complete, i.e. it is not defined for every possible expression in the
%% right hand side of a program rule. Basically the problem appears when
%% function composition is used. For those cases weakening can be applied
%% by replacing any term occurring in a right hand side of the fusioned
%% form of $\pi_2 \circ f$ by a term denotationally greater.
%% The simplest way of weakening is to replace the offending term by $\mathbf{1}$.
%% This is the solution we are using in our automated analizer. Of course,
%% this solution is safe, but work in progress indicates that a lot
%% can be done by using equational laws.

\section{From Partial Predicates to Set Expressions}\label{dac}
The most natural interpretation of a partial predicate 
$\pi \in \ppty{\alpha}$ is a subset of the domain $D_\alpha$, a set of
trees, more exactly an \emph{ideal} set of \emph{partial} trees.  
This section is devoted to show that set expressions and set
constraint based analysis \cite{papo:1997:acprc} 
can be used as a framework for the checking and inference of partial
predicate typings.

\subsection{Basic Notions}
\emph{`Set constraints are first-order logic formulae interpreted over the
domain of sets of trees'} \cite{papo:1997:acprc}.  
Set expressions ($e$) are expressions built from variables 
interpreted over sets of trees, function symbols intepreted as
functions over sets of trees and standard set operators (union,
intersection, inclusion, complement, etc.).   
A system of set constraints is a conjunction of inclusions of the form
$e_l\subseteq e_r$ with some restrictions on the set expressions
that can appear in the left or right hand sides.

\emph{Co-definite} set constraints \cite{chpo:1998:csc} is the class
of set constraints where constraints are inclusions between
\emph{positive} set expressions and where the set expression in the
left hand side is restricted to contain variables, constants, unary
function symbols and the union operator.\footnote{In
  Definition~\ref{def:codefinite} a restricted but equivalent
  characterisation is proposed.}  
Sets of co-definite constraints, if
satisfiable, always have a greatest solution.  The satisfiabilty
problem for co-definite set constraints is DEXPTIME-complete and an
algorithm is given in \cite{chpo:1998:csc}.

The restriction to positive expressions -- i.e.\ without the use of
complementation -- is essential for our purposes, as the complement of an
ideal is not an ideal.  The rest of operators are continuous, so we can
translate existing results in set constraint theory to our domains.
%% UFFF, esto hay que desarrollarlo mucho más.

The following definitions formalize these notions.

\newcommand{\VAR}{\mathbf{VS}}
\newcommand{\UNION}{\cup}
\newcommand{\PROJ}[2]{#1_{(#2)}^{-1}}

\begin{definition}[Set Expressions]
  Given a typed alphabet $\Sigma$ with constants ($a, b, c, \dots$)
  and nonconstant function symbols ($f$, $g$, $h$, \dots) and a typed
  set $\VAR$ of variable symbols ($u$, $v$, $x$, $y$, \dots), \emph{set
    expressions} follow the syntax:\footnote{The symbol originally used
    in set constraint theory is $\bot$.}
  \begin{eqnarray*}
    e & ::= & x ~|~ a ~|~ f(u_1,\dots,u_n) ~|~ \PROJ f k (u) 
                ~|~ e_1 \UNION e_2 ~|~ \emptyset \,.
  \end{eqnarray*}
  This syntax represents finite and infinite trees and we will use the
  notation $T_\Sigma$ for the whole set of well-formed (w.r.t.~types)
  trees.
\end{definition}

\begin{definition}[Valuation]
  A \emph{valuation} ($\sigma$) is a function from variable symbols to
  proper subsets of $T_\Sigma$ ($\sigma : \VAR \rightarrow
  2^{T_\Sigma}$).
\end{definition}

\begin{definition}[Interpretation of Set Expressions]
  Given a valuation $\sigma$, the \emph{standard interpretation}
  $I_\sigma$ of set expressions over $\Sigma$ and $\VAR$ is defined
  as\footnote{Where symbols $\cup$ and $\emptyset$ are overloaded.}
  \begin{eqnarray*}
    I_\sigma (x) & = & \sigma (x)\\
    I_\sigma (a) & = & \{a\}\\
    I_\sigma (f(u_1,\dots,u_n)) & = & \{f(t_i,\dots,t_n) ~|~ \forall i \in \{1,\dots,n\}.~ t_i \in I_\sigma(u_i)\}\\
    I_\sigma (\PROJ f k (u)) & = & \{t ~|~ \exists t_1,\dots,t_n. ~t_k = t ~\wedge~ f(t_1,\dots,t_n) \in I_\sigma(u)\}\\
    I_\sigma (e_1 \UNION e_2) & = & I_\sigma(e_1) \cup I_\sigma(e_2)\\
    I_\sigma (\emptyset) & = & \emptyset \,.
  \end{eqnarray*}
\end{definition}

\begin{definition}[Solution]
  A valuation $\sigma$ is a \emph{solution} of a set constraint %
  $e_l \subseteq e_r$ iff
  \begin{displaymath}
    I_\sigma (e_l) \subseteq I_\sigma (e_r) \,.
  \end{displaymath}
\end{definition}

\begin{definition}[Satisfaction]
  A system of set constraints $S$ is \emph{satisfiable} if there is
  some valuation $\sigma$ that is a solution of every constraint in
  $S$.
\end{definition}

\begin{definition}[Co-definite Set Constraints]
  \label{def:codefinite}
  A constraint $\varphi$ is a co-definite set constraint when it follows
  the syntax:
  \begin{eqnarray*}
    \tau & ::= & x ~|~ f(u_1,\dots,u_n) ~|~ \tau_1 \cup \tau_2 ~|~ \emptyset\\
    \varphi & ::= & a \subseteq x ~|~ x \subseteq \tau ~|~ x \subseteq f_{(k)}^{-1}(u) \,.
  \end{eqnarray*}
  We will use the notation %
  $\{\varphi_1, \varphi_2, \dots, \varphi_n\}$ to refer to the
  system %
  $\varphi_1\wedge\varphi_2\wedge\dots\wedge\varphi_n$.
\end{definition}

\subsection{Co-definite Set Constraints and Partial Predicates}
A partial predicate $\pi$ is interpreted as the set %
$\pi^{-1}(\True) = \{x \in T_\Sigma ~|~ \pi(x) = \True\}$ that is in
the codomain of interpretations of set expressions.  
Conversely, if $S$ is an ideal, $\Pi(S)$ will denote its corresponding
partial predicate, i.e.~$\Pi(S)(x)=True$ if $x\in S$, otherwise
$\Pi(S)(x)=\bot$. 
We will encode
partial predicates as variables and the greatest solution of a system
of co-definite set constraints.

\begin{example}[Some basic partial predicates]
  The following table shows how some partial predicates can be encoded
  ($z$, $s$, $n$ and $c$ refer, respectively, to constructor predicate
  transformers $\ppzero$, $\ppsucc$, $\ppnil$ and $\ppcons$ as described
  in Definition~\ref{def:ppconstructors}):
  \begin{center}
    \begin{tabular}{ccc}
      \textbf{Partial predicate} &
      \textbf{System of set constraints} &
      \textbf{Variable}\\
      $\pphnfnat = z ~\pplub~ s(\ppany)$ &
      $\{\mathit{hnf} \subseteq x \UNION y, x \subseteq \mathit{Zero}, y \subseteq \mathit{Succ}(\_)\}$ &
      $\mathit{hnf}$\\
      $\ppnfnat = z ~\pplub~ s(\ppnfnat)$ &
      $\{\mathit{nf} \subseteq x \UNION y, x \subseteq \mathit{Zero}, y \subseteq \mathit{Succ}(\mathit{nf})\}$ &
      $\mathit{nf}$\\
      $\ppnflistnat = n ~\pplub~ c(\ppnfnat,\ppnflistnat)$ &
      $\{\mathit{nf} \subseteq u \UNION v, u \subseteq [], v \subseteq (\mathit{nf}':\mathit{nf}),$ &
      $\mathit{nf}$\\
      &
      $\mathit{nf}' \subseteq x \UNION y,  x \subseteq \mathit{Zero}, y \subseteq \mathit{Succ}(\mathit{nf}')\}$ &
      \\
      $\ppspine = n ~\pplub~ c(\ppany,\ppspine)$ &
      $\{\mathit{snf} \subseteq u \UNION v, u \subseteq [], v \subseteq (\_:\mathit{snf})\}$ &
      $\mathit{snf}$
    \end{tabular}
  \end{center}
  where $\_$ represents fresh variables.
\end{example}

\begin{example}[Dependency]
  The intrinsic difficulty of working with dependency that was already
  patent in~\cite{MHM93} can be put in relation with the syntax of set
  constraints.  A typical property that can be represented with
  dependent demand patterns is that a pair is made of lists of the
  same length.  For instance, a demand typing for function
  $\fname{zip}$
\begin{verbatim}
zip :: ([a],[b]) -> [(a,b)]
zip ([],[]) = [];                   zip (x:xs,y:ys) = (x,y) : (zip (xs,ys))
\end{verbatim}
  is $\fname{zip} ~:~ \ppspine \need samelength$ where
\begin{verbatim}
samelength :: PP ([a],[b])
samelength ([],[]) = True;    samelength (x:xs,y:ys) = samelength (xs,ys)
\end{verbatim}
  The greatest solution to the following set constraint
  system for the variable $sl$ captures the dependent information of
  the partial predicate $\samelength$:
  \begin{displaymath}
    \{sl \subseteq (l_1,l_2) \UNION (l_1',l_2'),\ %
    l_1 \subseteq [],\ %
    l_2 \subseteq [],\ %
    l_1' \subseteq x:xs,\ %
    l_2' \subseteq y:ys,\ %
    (xs,ys) \subseteq sl\} \,.
  \end{displaymath}
  But this system is not co-definite (last constraint has a binary
  function symbol in the left hand side).  In order to get a
  co-definite set constraint system, the last constraint is
  substituted by two constraints: %
  $xs \subseteq \PROJ {(,)} 1 (sl)$ and %
  $ys \subseteq \PROJ {(,)} 2 (sl)$.  With the substitution we
  have lost the dependency information.  Nevertheless, the solution is
  correct with respect to the interpretation of the partial predicate
  in the following formal sense:
  \begin{displaymath}
    \samelength^{-1}(\True) \subseteq I_\sigma(sl)
  \end{displaymath}
  where $\sigma$ is the greatest solution to the system of co-definite
  set constraints.
\end{example}

\subsection{Generating Co-definite Set  Constraints}
DAC (\emph{Demandedness Analysis for Curry}) is a tool that generates
a system of co-definite set constraints from a given program.  In this
section we explain how DAC generates the system.  Observe that the
solving of a system of set constraints is completely independent of
the application, i.e.\ the fact that we are encoding partial
predicates is immaterial. 
% hmmm?
Observe, as well, there could be other ways to generate correct
systems of constraints.

In the first place, we need to introduce the logic that relates the
variables in the system of set constraints with the meaning of the
program.  This connection relies on the fact the partial predicates
the user is interested in are defined as functions in the kernel
language.

\newcommand{\oo}{{\small \bullet\!}}

\begin{definition}[Variable Construction]
  Given $f \in \mathit{FS}$ and $p \in \mathit{FS}$, with types
  $\tau_1 \fl \tau_2$ and $\PP\ \tau_2$, respectively, the infix
  operator $(\oo\,)$ is used to construct a new variable %
  $p\oo f \in \VAR$ that represents the degree of evaluation
  demanded by $f$ in order to give a result as evaluated as $p$.

  Auxiliary variables are introduced for different subterms in the
  program equations that define $f$:
  \begin{itemize}
  \item[--] $p\oo f.i$ refers to the demandedness information introduced
    by the $i$-th equation defining $f$
%  \item[--] $p\oo f.i.1$ refers to the left hand side of the $i$-th
%    equation defining $f$.
  \item[--] $p\oo f.i.pos$ refers to the subterm $lhs|_{pos}$ if $lhs$
    is the left hand side of the $i$-th equation defining~$f$.
  \item[--] $p\_f.i.pos$ refers to the subterm $rhs|_{pos}$ if $rhs$
    is the right hand side of the $i$-th equation defining $f$.
  \end{itemize}
  Finally, the constraint generation algorithm generates variables of
  the form $(q\oo g)\oo f$.  Although these can be given a neat
  interpretation, the constraint solver will treat them as
  indivisible, so they will have to be transformed in some way in
  order to be useful.
\end{definition}
The intuitive meaning of $d\oo f$ is to denote (an approximation
of) $d\circ f\,$.\footnote{%
  Hence the choice of the symbol `$\!\oo\,$'.} 
This connection will be formalized below.
The intuitive meaning of $p\oo f.i.pos$ is the projection at
position $pos$ of the set $p\oo f.i\,$.
Variables generated from positions in the right hand sides have a less
evident meaning or, perhaps, more operational but they provide
valuable information for compilation.

\newcommand{\NF}{\textrm{nf}}
\newcommand{\LT}{\subseteq}

\begin{figure*}
\begin{eqnarray}
\NF\oo\plus & \LT & \NF\oo\plus.1 ~\UNION~ \NF\oo\plus.2\\
\NF\oo\plus.1 & \LT & (\NF\oo\plus.1.1, \NF\oo\plus.1.2)\\
\NF\oo\plus.1.1 & \LT & \Zero\\
\NF\oo\plus.1.2 & \LT & \NF\\
\NF\oo\plus.2 & \LT & (\Succ(\NF\oo\plus2.1.1), \NF\oo\plus.2.2)\\
\Succ(\NF\_ \plus.2.1) & \LT & \NF\\
%% OJO!! mirad esto:
%(\NF\oo\plus.2.1.1, \NF\oo\plus.2.2)) &\LT & \NF\_ \plus.2.1
\NF\oo\plus.2.1.1 &\LT& \NF\_ \plus.2.1.1.1 \\
\NF\oo\plus.2.2   &\LT& \NF\_ \plus.2.1.1.2 \\
(\NF\_ \plus.2.1.1.1, \NF\_ \plus.2.1.1.2)) &\LT & \NF\oo\plus
\end{eqnarray}
% \begin{align*}
% \NF\oo\plus &\LT 
%     (\Zero, \NF) \cup 
%     (\NF\oo\plus.2.0.1, \NF\oo\plus.2.0.2) \\
% \NF\oo\plus.2.0.1 &\LT \Succ(\NF\oo\plus.2.0.1.1) \\
% (\NF\oo\plus.2.B.0.1, \NF\oo\plus.2.B.0.2) &\LT \NF\oo\plus\\
% \NF\oo\plus.2.0.1.1 &\LT \NF\oo\plus.2.B.0.1\\
% \Succ(\NF\oo\plus.2.B.0.2.1) &\LT \NF\oo\plus.2.B.0.2\\
% \NF\oo\plus.2.0.2 &\LT \NF\oo\plus.2.B.0.2.1\\
% \end{align*}
%
\caption{Set of constraints generated from program \textit{plus}}
\label{fig:dacplus}
\end{figure*}

\begin{example}
  Figure~\ref{fig:dacplus} shows a system of constraints generated
  from function \textit{plus} (Example~\ref{ex:plus}) to give a result
  in normal form.
\end{example}

\subsection{Generating Systems of Set Constraints}\label{reglas}
The generation scheme is presented here as a set of rules. These will
be stated in a moderately informal way, in order to hide some of the
details to the reader, especially those concerned with the
handling of occurrence indices.

\begin{myrule}[Main Function Constraint]
  Given a partial predicate symbol $p$ and function symbol $f$, with
  defining rules
  \begin{eqnarray*}
    f ~t_1 = b_1 
    ~~~~\dots~~~~ 
    f~t_n = b_n 
  \end{eqnarray*}
  the following constraint is generated:
  \begin{eqnarray*}
    p\oo f & \LT & p\oo f.1 ~\UNION~ \dots ~\UNION~ p\oo f.n \,.
  \end{eqnarray*}
\end{myrule}

\begin{myrule}[Main Rule Constraint]
  For every rule 
  \begin{eqnarray*}
    f_i~(t_1, \dots, t_n) &=& b_i
  \end{eqnarray*}
  the following constraint is added:
  \begin{eqnarray*}
    p\oo f.i \LT (p\oo f.i.1, \dots, p\oo f.i.n) \,.
  \end{eqnarray*}
  Notice that this is a \emph{lossy} step, i.e.\ possible dependencies
  among the arguments through the body of the rule $(b_i)$ can be
  lost. This means that an analyzer based on program transformation
  techniques can, at least theoretically, achieve a better accuracy.
\end{myrule}

\begin{myrule}[Head Constraints]
  For every rule 
  \begin{eqnarray*}
    f_i~(t_1, \dots, t_n) &=& b_i
  \end{eqnarray*}
  and for every $j \in \{1,\dots,n\}$ the following constraint is
  added:
  \begin{eqnarray*}
    p\oo f.i.j &\LT& \Delta(p,f,i.j,t_j) \,,
  \end{eqnarray*}
  where $\Delta$ is the function that constructs a set expression from
  a term by replacing every occurrence of a program variable with a
  demand variable decorated with its position, i.e.:
  \begin{eqnarray*}
  \Delta(p,f,i,c(t_1,\dots,t_m)) &=& 
    c(\Delta(p,f,i.1,t_1),\dots,\Delta(p,f,i.m,t_m)) \\
  \Delta(p,f,i,x)                 &=&
    p\oo f.i \,.
  \end{eqnarray*}
  This step usually
  generates superfluous constraints of the form $v \LT v$ which can be
  discarded later.
\end{myrule}

\begin{myrule}[Body Constraints]
  We can distinguish several cases here:
  \begin{enumerate}
  \item (\emph{The body is a variable}) If the rule is of the form
    \begin{eqnarray*}
      f_i~(t_1, \dots, t_n) &=& x
    \end{eqnarray*}
    $x$ being a program variable, the constraint
    \begin{eqnarray*}
      p\oo f.i.1.pos & \LT & p
    \end{eqnarray*}
    where $pos$ is the position where $x$ occurs in the left hand side,
    is added to the system.
  \item (\emph{The body is a constant}) If the rule is of the form
    \begin{eqnarray*}
      f_i~(t_1, \dots, t_n) &=& k
    \end{eqnarray*}
    $k$ being a constant, the constraint
    \begin{eqnarray*}
      k &\LT& p
    \end{eqnarray*}
    is added to the system. This constraint will often be trivial.
  \item (\emph{The body is a function application}) This is the
    clumsiest case. To simplify the presentation, let us assume,
    without loss of generality, that the form of the rule is the
    following: 
    \begin{eqnarray*}
      f_i~(\bar{t}) = g~(h_1(\bar{t}), \dots, h_m(\bar{t})) \,.
    \end{eqnarray*}
    The constraints
    \begin{eqnarray*}
      p\_f.i.1 &\LT& \PROJ{(,)} 1 {(p\oo g)} \\
               &\vdots&         \\
      p\_f.i.m &\LT& \PROJ{(,)} m {(p\oo g)}        
    \end{eqnarray*}
    are added to the system, and also the constraints :
    \begin{eqnarray*}
      p\oo f.i &\LT& p\_ f.i.1\oo h_1 \\
      &\vdots& \\
      p\oo f.i &\LT& p\_ f.i.m\oo h_m \,.
    \end{eqnarray*}
    Notice that this step is responsible for the appearance of
    `nested' demand variables.
  \end{enumerate}
\end{myrule}

\begin{myrule}[Simplification]
In this step two kind of actions are performed: the shortcut of
transitive chains and the simplification of nested demand variables.

If a variable of the form $(p\oo g)\oo f$ is found, and a concrete
representation $p'$ for $(p\oo g)$ is known -- which is usually the
case when $g$ is a data constructor -- then it is replaced by $p'\oo
f$.  This step is necessary when standard -- i.e.\ \emph{problem
  independent} -- techniques for solving the constraint systems are
going to be used.
\end{myrule}

\begin{myrule}[Weakening]
  Sometimes it is not easy to compute the $(p\oo g)$ of the previous
  step, so some sort of approximation is necessary, i.e.\ using any
  $p''$ satisfying $p'' \sqsupseteq p'$ instead of $p'$.  This is
  often possible.  In the worst case $\ppany$ can be used.
\end{myrule}

The following result states the soundness of the analysis based on the
solution of this set of constraints.

\begin{theorem}\label{soundness}\emph{(Soundness of the Analysis)}
Let $S$ denote the system of constraints generated from a given
program applying the rules above. Let $\sigma$ be a solution of
$S$. For every variable $d\oo f$ occurring in $S$ the following must
hold:
\[ \Pi\syntb{\sigma(d\oo f)} \sqsupseteq d\circ f \,.\]
\end{theorem}
We will just sketch the proof here. A more detailed explanation can be
found in \ref{app:proof}.
%% OJO!!
The idea is to apply a set of program transformation rules 
to $d\circ f$, for every possible
combination of $d$ and $f$, so that the resulting program is
structurally similar to $S$.

%(This was Theorem~\ref{soundness})
%%%%%%%%%%%%%%%%%%%%%%%%%%%%%%%%%%
\section{Application to Code Generation}\label{codegen}
For the sake of brevity, we will not develop the issues related to
code generation in full here. 
A detailed discussion can be found in \cite{mitesis}, first in an
abstract fashion -- by means of an operational semantics driven by
degrees of definiteness --  and then in the context of a stack-based machine. 
Anyway, code generation from the demand information (represented by
partial predicates) is a challenging
task on its own and some of the details of a full
compiler are still open.

%% % In Subsection~\ref{lazynar}, we have explained which properties an
%% % efficient and lazy implementation, and hence the translation scheme,
%% % should have.
%% Let us first state the desired properties of the
%% translation
%% %H
%% scheme:
%% %
%% \begin{itemize}

%% \item it should evaluate as much demanded arguments as
%%   safely possible before the function call in order to avoid
%%   reevaluations,

%% \item when backtracking it should first try the next possible
%%   rule for the considered function before trying alternative solutions
%%   for the arguments,

%% \item it should be as lazy as the naive lazy strategy, i.e.\  it
%%   should only evaluate an expression, if the naive lazy strategy would
%%   eventually also do so,

%% \item it should take into account dependencies between the
%%   arguments.

%% \end{itemize}
%% %
%% Most of these properties are guaranteed by an appropriate use of
%% the inferred demand patterns.

The basic idea is that different, specialized versions of a given function can
be compiled for different degrees of evaluation demanded on its result.
For instance, if the main goal of a certain program is
%
%\begin{small}
\begin{verbatim}
?- mergeSort (f x)
\end{verbatim}
%\end{small}
%
\noindent
the result must be shown in normal form, so a special version
\textit{mergeSort\_nf} will be generated. In order to give a result in normal
form, the argument to \textit{mergeSort} must also be a total value, which
implies that $f$ can also be replaced by an specialized version --
$f\_\mathit{nf}$ -- and so on, i.e.\ demand is back-propagated from
the result to the argument expressions.

\paragraph{Some Experiments}\label{results} %H
The following example programs were executed on a stack-based
narrowing machine~\cite{lbam}. %H
The narrowing machine is an extension of a purely functional
machine enriched by mechanisms for unification and
backtracking, similar to the WAM.

Based on the implementation of the presented ideas on the stack-based
narrowing machine, we have tried some example programs and measured
their runtimes with the naive lazy approach and with our new
approach. Additionally, we have measured the runtimes for eager
narrowing. %H

We have investigated the following example programs: 1) the
computation of all the sublists of %H
\textit{reverse} $[1,\dots,n]$ , and 2) the $n$-queens problem %H
(using a simple generate and test approach). 
Both examples have the property that a lot of reevaluations
are needed since demanded arguments are not evaluated in advance.
Due to space limitation we omit the code.
The runtimes are depicted in Table~\ref{expres}.
\begin{table}
\begin{center}
\begin{tabular}{rrrrr}
\hline
\emph{example} & $n$ & \emph{eager} & \emph{naïve lazy} & \emph{demand
driven} \\
\hline
         & 10 &  0.47 &  1.00 & 0.62 \\%H
         & 11 &  0.91 &  1.98 & 1.24 \\%H
\emph{sublists}
         & 12 &  1.86 &  3.96 & 2.42 \\%H
         & 13 &  3.68 &  7.94 & 4.80 \\%H
         & 14 &  7.42 & 15.82 & 9.60 \\%H
         & 15 & 15.05 & 31.75 & 19.23 \\%H
\hline
         & 4 &   1.06 &   0.89 &  0.21 \\%H
\emph{$n$ queens}
         & 5 &  14.50 &  10.55 &  2.60 \\%H
         & 6 & 247.03 & 174.30 & 39.30 \\%H
\hline
\end{tabular}
\end{center}
\caption{\label{expres}Runtimes of the example programs in
seconds.}
\end{table}
The examples show that the runtimes can be considerably improved,
if the demanded
arguments are evaluated in advance. Notice that in examples like
$n$-queens, the lazy strategy is even better than the eager one.
Moreover, bigger real examples have a lot of nested function
calls, which implies a considerable risk of reevaluation.
%

%% Resultados en el artículo con ajimenez
Table~\ref{sublists-prolog} shows the results obtained with the
$\mathit{sublists}\circ\mathit{reverse}$ example using the translation into
Prolog, with and without code optimization based on demand analysis.
The measures have been taken in discrete resolution steps.

\begin{table}
\begin{center}
\begin{tabular}{rrrrl}
\hline
$\mathit{sublists}\circ\mathit{reverse}~[1,\dots,n]$ & $n$ & \emph{without
demand anal.} & \emph{with demand anal.}  & \emph{ratio} \\
\hline
& 3 & 41& 36& 1.13\\
& 4 & 88        & 72    & 1.22\\
& 5 & 183               & 141   & 1.29\\
& 6 & 374               & 275   & 1.36\\
& 7 & 757       & 538   & 1.40\\
& 8 & 1524      & 1058  & 1.44 \\
\hline
\end{tabular}
\end{center}
\caption{Results in a lazy \emph{producer-consumer} scheme.}
\label{sublists-prolog}
\end{table}

\section{Related Work}\label{related}
Our original work  on demand analysis~\cite{JMM92,MKM93,HM93,MHM93}
was based on the generation and solution of a set of \emph{demand
  equations} that were solved in a domain of regular trees
(\emph{demand patterns}). Similar, in spirit, to the techniques
presented in Section~\ref{dac}, a semantic justification was missing
and the solving method was ad-hoc. Partial predicates provide the
necessary semantic ground and the advances in set constraint resolution
makes unnecessary to reinvent the wheel.

With respect to partial predicates, the most striking similarity is
with \emph{projection analysis}~\cite{wadlerhughes87projections}. 
However, the rationale and
meaning for these two formalisms differ in some key aspects. A
\emph{projection}, in a domain-theoretic sense, is an idempotent
approximation to the identity (in a given type), i.e.~%
$\alpha::\tau \fl \tau$ is a projection (in $\tau$) iff 
$\alpha \menor id$ and $\alpha \circ \alpha = \alpha$.

While partial predicates try to be an extension of classic strictness
analysis, projection analysis are designed to capture the property
that a given function is invariant under certain program
transformations. The typical example is \emph{head-strictness}, the
property that a function on lists gives the same results when the list
constructor in its argument is replaced by a version strict in its
first argument. Mathematically, this transformation is a projection 
$\HP::[a]\fl[a]$, so the property of $f$ being head-strict is expressed
as $f = f\circ\HP$. In general, projection analysis studies properties of the form 
$\alpha\circ f = \alpha\circ f\circ\beta$, where $\alpha$ and $\beta$
are projections. These are abbreviated as $f:\alpha\Rightarrow\beta$.

Properties expressible in both formalisms are different.
First of all, let us show that head strictness cannot be represented
by a partial predicate typing.

\begin{theorem}
There is no pair of partial predicates $\pi_1, \pi_2$ such that the
set of functions $\{f|f:\pi_1\need\pi_2\}$ coincides with that of the
head-strict ones. 
\end{theorem}
\begin{proof}
Let us note that the property of being head-strict is just `too
polymorphic' as it does not take into account the type of the result,
so equivalence just makes sense fixing a particular type,
i.e.~considering just the head-strict functions for a given type 
$[\sigma]\fl\tau$. This makes the proof shorter, as we can restrict
ourselves to the type $[Bool]\fl Bool$. There will be just five
possibilities for $\pi_2$: $nothing$, $true$, $\false$, $\hnf$ and
$any$. The key to the proof is in considering the functions $any$,
$nothing$ (which are head-strict) and $\spine$ (which is not). 
Any combination of partial predicates which would hold for both $any$
and $nothing$ would also hold for $\spine$, contradiction.
\end{proof}

Projections, on the other hand, are able to express partial predicate
typings, but only if a tricky artifact is added to the formalism:
assuming the existence in the domain of a new element ($\abort$) less
defined than $\bot$. This had to be introduced by Wadler and Hughes in
order to capture classic strictness with projections, but complicates
the formalism in several ways. The following result holds assuming
programs are $\abort$-strict:

\begin{theorem}
For every pair of partial predicates $\pi_1, \pi_2$ there is a pair of
projections $\alpha, \beta$ such that the set
$\{f|f:\pi_1\need\pi_2\}$ coincides with 
$\{f|f:\alpha\Rightarrow\beta\}$ 
--- under reasonable type restrictions.
\end{theorem}
\begin{proof*}
The proof is constructive. Define $\alpha$ and $\beta$ in the
following way: 
\[
\begin{array}{l@{\;}lr@{~~~~~~~~~~~~~~~~~~}l@{\;}lr}
\alpha~x      &= x      & \mathrm{if}~~\pi_2~(f~x) = true &
\beta~x       &= x      & \mathrm{if}~~\pi_1~x = true \\
\alpha~x      &= \abort & \mathrm{otherwise}          &
\beta~x       &= \abort & \mathrm{otherwise}          \,.%%\\
%%\alpha~x      &= \bot   & \mathrm{otherwise}\\
\end{array}
\]
Let us examine both implications:
\begin{enumerate}
\item[(i)] ($f:\pi_1\need\pi_2 \implies f:\alpha\Rightarrow\beta$)\\
  There are two possibilities for any argument $x$ to $f$:
  \begin{enumerate}
  \item[a)] ($x \in\pi_1$) Trivial:
            $(\alpha\circ f\circ\beta)~x = (\alpha\circ f)(\beta~x) =
             (\alpha\circ f)~x$.
  \item[b)] ($x \notin\pi_1$) In this case we know that 
            $f~x \notin\pi_2$. So, in one hand we have:
            \[ (\alpha\circ f\circ\beta)~x =
               \alpha(f(\beta~x)) = 
               \alpha(f~\abort) =
               \alpha~\abort =
               \abort \,.\] 
            On the other hand, using the fact that $f~x \notin\pi_2$,
            $\alpha(f~x) = \abort$.
  \end{enumerate}
\item[(ii)] ($f:\alpha\Rightarrow\beta \implies
                  f:\pi_1\need\pi_2$)\\
           Using \emph{reductio ad absurdum}: suppose there is some
                  $z$ s.t.\ $z\notin\pi_1$ and $f~z\in\pi_2$.
           Then it is trivial to show that 
           $(\alpha\circ f\circ\beta)~z = \abort$ and
           $(\alpha\circ f)~z = f~z$, contradiction.~~~~~\proofbox
\end{enumerate}

\end{proof*}

Projections in the lifted domain are no longer expressible in source
code, precluding the possibility of using program transformation or
the other techniques that are applicable to partial predicate typings.

%Revisar
The use of program transformation techniques for program analysis
is used in other approaches, like \emph{abstract compilation},
%References
but to our best knowledge, the application of the fold/unfold method
of program transformation for program analysis is a novel
idea. The only similar approach appears in \cite{Gallagher2000}
to develop a type inference system for Prolog, and
\cite{CominiGoriLevi2000} to verify program properties.

There is also some existing work on using constraint generation for
this kind of problems. In~\cite{sekar95fast}, two degrees of
definiteness are defined: normal form and head normal form, which
leads to the notion of $e$-demand (normal form needed) and $d$-demand
(head normal form needed). Recursive equations for computing how these
degrees of demand are propagated are generated for a given program,
based on an operational semantics. The authors also mention the
possibility of using demand analysis for sequentiality recovery,
although the idea is not developed there.

\section{Open Issues and Future Work}\label{future}
%% We have yet studied how the most used strictness analyses can be
%% understood in our framework. The uniform presentation allows
%% to compare the accuracy of the analyses and it is clear that
%% demand analysis provides more information than the other ones.

The importance of demandedness analysis goes beyond functional-logic
languages. In \cite{FalaschiHW00} dependent demand patterns are used for
the analysis of concurrent (constraint) logic languages.

%% Partial evaluators for Curry are currently being developed, see for instance
%% \cite{ramos:silva:vidal:2005:icfp}. 
%% We plan to use them in the short term, both as an
%% experimentation of this program transformation approach to program analysis
%% and also as a way of benchmarking the evaluators themselves.

%% Our use of the fold/unfold method of program transformation was not intended
%% as a way of implementing the analysis in the first place, but after using
%% it as a proof assistant we are inclined to think that partial evaluators can
%% actually be used as analyzers when other approaches, like abstract
%% interpretation or set constraints do not produce acceptable
%% information.

%% Transformación de programas: ¿viable?
The question whether program transformation tools can be used for
program analysis following the techniques presented here is still
open, although several problems appear. In the first place, deciding
on the equality of functions is harder, in general, than checking the
equality of set expressions. Second, although the theory behind
program transformation is well developed, practical implementations
are scarce. However, this is an active area and we plan to study the
possibility of adapting the tools by Vidal's group%
~\cite{ramos:silva:vidal:2005:icfp} to serve this purpose.

Another possible extension of this work is to study the application of
partial predicates to other analysis problems, like groundness, etc.

The extension of the analysis when higher order functions are used
deserves an additional discussion.
In fact, some complications do appear if higher order definitions are
introduced. Let us consider an example involving
curried definitions.

\noindent
Take, for instance, the standard definition of the addition of Peano
naturals of example \ref{ex:plus}.

%\begin{small}
%\begin{tabbing}
%%\texttt{~~~~}
%\=\kill
%\>\verb.(+) :: Nat -> Nat -> Nat.\\
%\>\verb.0     + m = m.\\
%\>\verb.(S n) + m = S (n + m).
%\end{tabbing}
%\end{small}

%\noindent
%Its noncurried version can be defined in this way:

%\begin{small}
%\begin{tabbing}
%\texttt{~~~~}
%\=\kill
%\>\verb.plus :: (Nat, Nat) -> Nat.\\
%\>\verb.plus (0,   m) = m.\\
%\>\verb.plus (S n, m) = S (plus (n, m)).
%\end{tabbing}
%%\end{small}

%\noindent
An interesting property that we would like to express is the fact that
the first argument must be evaluated to head normal form in order to get
a result in head normal form. This is very simple for the noncurried form:
\( \plus ~:~ (\pphnfnat \times \ppany) \need \pphnfnat\),
but it is not clear at all how to express that for the curried version.
In first place, \texttt{(+)} maps naturals to \emph{a new function}, and it is
not this function we are interested in, but the result of applying it to any
other natural number. To grab the problem more formally, we will make use of
the following lemma.

\begin{lemma}[Currying lemma] 
Let $(f\,\circ)$ denote $\lambda\,x.\,f \circ x$. Then, the following
holds:
\[ \mathrm{curry}~(f \circ g)~ = ~ (f\, \circ) \circ
     (\mathrm{curry}~g) \,.
\]
%%\noindent
%%\textbf{Proof:} Trivial rewriting. 
%\noindent$\Box$
\end{lemma}
What we are looking for is a property of the form:
%\begin{displaymath}
$
  \texttt{(+)} : \pphnfnat \need \pi
$
%\end{displaymath}
and what we actually have is
%\begin{displaymath}
$
  \plus : (\pphnfnat \times \ppany) \need \pphnfnat \,.
$
%\end{displaymath}
This is equivalent to:
\begin{displaymath}
  \pphnfnat \circ \plus \sqsubseteq (\pphnfnat \times \ppany) \,.
\end{displaymath}
As $\mathit{curry}$ is continuous, we can curry both sides of the inequality:
\begin{displaymath}
  \fname{curry}~(\pphnfnat \circ \plus) \sqsubseteq \fname{curry}~(\pphnfnat \times \ppany)
\end{displaymath}
and using the lemma above:
\begin{displaymath}
  (\pphnfnat\,\circ) \circ \texttt{(+)} \sqsubseteq \fname{curry}~(\pphnfnat \times \ppany)
\end{displaymath}
or equivalently
\begin{displaymath}
  \texttt{(+)} : \fname{curry}~(\pphnfnat \times \ppany) \need (\pphnfnat\,\circ) \,.
\end{displaymath}

Well, this gives \emph{essentially} the same information as the demand 
typing for the noncurried version, but there is a problem: the functions 
in the typing are no longer partial predicates, i.e.\ they are not in the
domain $\tau \fl \Two$.

There are essentially two ways of dealing with these problems in practice.
One possibility is to avoid higher order definitions as much as possible,
translating curried versions to noncurried ones -- and vice-versa with the
results of the analysis.

The other possibility is to generalize demand types to pairs of functions
in the domain $\tau \fl \tau'$. Although the theoretical interest of
this lifting to higher order domains is apparent, and a higher order 
metalanguage seems feasible, the practical use will be very restricted,
as the techniques in Section~\ref{sec:inference} will not be applicable.

\section{Conclusion}\label{demand:conclusion}

We have presented a semantic framework for the denotation of 
demand properties, decoupling it from its abstraction or 
any implementation detail of the analysis. 
In spite of defining a new language and an associated demand logic
(cf.~the strictness logic in~\cite{benton:phdthesis}),
properties are expressed in the language under study and problem
independent techniques are used as proof methods: equational reasoning
or set constraint solving.

The collection
of analysis that can be modelled includes classic
strictness analysis, Wadler's four point domain, etc.
In particular it
allows to describe demand analysis that is very
important for the efficient implementation of
several aspects of functional-logic languages:
efficient implementation of lazy narrowing
\cite{MKM93,MHM93}, compilation of nonsequential programs
\cite{padl00} or the lazy management of
default rules \cite{lazydef}. 
The formalism has been used to prove the correctness of a method
based on set constraint solving, in a constructive way.
It can also be used to generate domains suitable for abstract interpretation 
(uniform predicates).

%% OBSCURE
%% The unified framework has some other advantages.
%% The most interesting is that different analyses
%% can be compared in terms of the amount of 
%% information provided and accuracy.

Furthermore, the formalism is quite intuitive as it
resembles the language to reason about and uses
program transformation techniques. 
We have also shown how polymorphism and higher-order properties can be
managed -- at least theoretically -- in this framework, although the
extension for the analysis of higher-order program presents some
challenges. 

An original feature of this research is that it is completely based
on a declarative (model theoretically) denotational semantics of the
language, rather than on an operational one.\footnote{%
  We include here denotational presentations of essentially
  operational semantics like~\cite{Zartmann:SAS97}.} 
From a practical point of view, performing the analysis on operational
data is often easier, but considering that the domain of properties
can be understood in a purely declarative setting, we wanted to
explore the possibility of performing the analysis without using a
particular operational semantics.

We also felt that the applications of set constraints to program
analysis have been biased towards problems stated in an operational
fashion, and that deriving set constraints from semantic equations was
an original approach and a challenge worth taking up.

\section*{Acknowledgements}
This research was supported in part by the Spanish MCYT grant 
TIC2003-01036. We also want to thank Enea Zaffanella and 
the anonymous referees for
their valuable comments on earlier versions of this paper.

\bibliographystyle{acmtrans}

%\bibliography{/home/xmc/bibtex/xmc,/home/xmc/bibtex/xeral}
%\bibliography{\bibdir/xeral,\bibdir/xmc,\bibdir/setconstraints}
\bibliography{\bibdir/babylon,\bibdir/publications}

%% A P E N D I C E S
\appendix
\clearpage

\section{Semantics of the Kernel Language}\label{app:semantics}
The following lines describe a denotational presentation of a
declarative semantics for the language used in this paper.
%% La declarativa antes
We start providing a declarative, logical semantics.
Let us define the semantic domains first. 
$H$ is the cpo completion of the Herbrand universe formed with
all the (data) constructors in a program. 
The (higher order) domain of values $D$ is given as the least solution
to the equation 
\[ D \cong H + [D_{\bot} \rightarrow D_{\bot}] + 
   \sum_{i} \{ (C~d_1 \ldots d_i)| C \in DC^i,~ \forall k.d_k \in D \} \]
Environments are type-preserving mappings from variable symbols to $H$,
and interpretations (for a given program) map every function
symbol to a value in $D$:
\begin{align*} 
  \Env         &= \VS \ra H \\
  \textit{Int} &= \textit{FS} \rightarrow D 
\end{align*}
Environments can be lifted in the standard way to functions from terms
(with variables) to $H$, with the usual overloading.
We regard constructors as free, and thus their denotation is the usual,
standard one.

\begin{definition} [Models]
An interpretation $I$ is a model of \emph{a ground instance} 
$l' = b' \ra r'$ of a defining rule $l = b \ra r$ iff 
\[ \valiun{l'} \mayor \valiun{b} \ra \valiun{r'} \]
An interpretation is a model of a rule when it models all its ground
instances:
\[ \forall \sigma.\ \valiun{\sigma l} \mayor 
      \valiun{\sigma b} \ra \valiun{\sigma r}   \]
being $\sigma$ a well typed grounding substitution.
An interpretation $I$ is a model of a program $P$ (in symbols
$I \models P$) iff $I$ is a model of every defining rule in $P$.
\end{definition}
\noindent 
Next we define a denotational construction for such a semantics.
We will make use of the following
semantic functions%
  \footnote{Properly speaking, they all depend on the
  program -- $\FF_P$, $\EE_P$, etc. -- but the subscript will be dropped
  when no confusion may arise.}: 
\[
        \begin{array}{l}
        \FF: \textit{Int}       \\
        \RR: \textit{Rule} \rightarrow \textit{Int} \rightarrow
                        \textit{Int} \\
        \EE: \textit{Exp}  \rightarrow \textit{Int} \rightarrow D
        \end{array}
\]
$\EE$ is just recursive evaluation of expressions according to the
semantics of the program, and can be defined as the homomorphic
extension of the semantics for function symbols ($\FF$).  
It is needed in order to evaluate the right hand sides of rules.
$\RR$ is the interpretation transformer associated with each rule of
the program and represents the amount of information added by every
possible application of that rule:
\begin{align*}
\RR \semb{f~t_1 \ldots t_n~\mathtt{=}~b \ra r} I      &=
  \lambda \mathit{fs}. (\mathit{fs} = f) \rightarrow
    \lambda x_1 \ldots x_n. \bigsqcup_{\rho \in \Env} 
      (\rho t_1 \seq x_1 \wedge \cdots \wedge \rho t_n \seq x_n
       \wedge \EE \semb{\rho b} I)
        \rightarrow \EE \semb{\rho r} I                                  \\
\FF_P &= 
  \mathit{lfp}(\bigsqcup_{\mathit{rule} \in P}(\RR \semb{\mathit{rule}}))
\end{align*}
The symbol ($\seq$) denotes strict equality and 
($\cdot \rightarrow \cdot$) is shorthand for 
($\cdot \rightarrow \cdot | \bot$). 
For every equational program $P$, $\FF_P$, $\RR$ and $\EE$ are
continuous\footnote{This is due to the operators involved in their
  definition. Observe that the \emph{lub} in the right hand side of
  the definition of \RR is not infinite because the conditional inside
  limits the possibilities to $\bot$ or $\EE \semb{\sigma r} i$
  -- where $\sigma$ is unique  -- and thus is well defined.}.
%
%% \begin{example}
%% The meaning of addition, as defined by the rules

%% %%      \begin{center}
%% %%      \begin{small}
%% %%      \begin{minipage}{3cm}
%% %%      \begin{tabbing}
%% %%      (s N) \=+ M \=:= s (N+M)        \kill
%% %%      0     \>+ M \>:= M.                     \\
%% %%      (s N) \>+ M \>:= s (N+M).
%% %%      \end{tabbing}
%% %\begin{small}
%% \begin{verbatim}
%%     0     + m = m
%%     (S n) + m = S (n + m)
%% \end{verbatim}
%% %\end{small}
%% %
%% %%      \end{minipage}
%% %%      \end{small}
%% %%      \end{center}
%% %
%% %\noindent 
%% is given by 
%% %
%% \[
%%         \FF \semb{+} = \lambda x. \lambda y. 
%%                 \textit{lfp} (\lambda i.
%%                         (x \seq 0 \rightarrow \EE \semb{y} i)
%%                         \sqcup
%%                         (\bigsqcup_{N,M}(x \seq (s~N) \rightarrow 
%%                                 \EE \semb{s(N+M)} i
%%                         ))
%%                 )
%% \]
%% after some simplification of the formula.               %\hfill $\Box$
%% \end{example}

\medskip\noindent
The following result, proved in \cite{mitesis}, states the adequacy of
both presentations:

\begin{theorem}
Let $\Rules_P(f)$ be the set of rules defining function symbol $f$ in
propgram $P$. 
For every functional-logic program $P$, 
and function symbol $f \in \FS_P$,
$\FF_P~f$ is the minimal model for the rules in $\Rules_P(f)\,$.
\end{theorem}

%%% Local Variables: 
%%% mode: latex
%%% TeX-master: "tplp-reviewed"
%%% End: 

\section{Proof of Theorem~\ref{soundness}}\label{app:proof}
Theorem~\ref{soundness} states the soundness of the analysis based on the
solution of a set of constraints.
If $S$ denotes the system of constraints generated from a given
program applying the rules in Subsection~\ref{reglas} and $\sigma$ is a solution of
$S$, then for every variable $d\oo f$ occurring in $S$ the following must
hold: 

The following is still very sketchy --- a full proof would be much longer. Some
lemmata on valid program transformations are necessary in order to
justify the different rules for constraint generation:

\begin{lemma}\label{lemma:transform}
The following program transformations are valid
according to the semantics of the kernel language:
\begin{enumerate}
\item Any function symbol $f$ defined by rules
\[
f~t_1 = b_1     
~~~~ \cdots ~~~~
f~t_n = b_n 
\]
can be rewritten as 
\[
f = f.1 
~~~~ \cdots ~~~~
f = f.n 
\]
where
\[
f.1~t_1 = b_1
~~~~ \cdots ~~~~
f.n~t_n = b_n 
\]

\item Any program rule
\[ f(t_1,\ldots,t_n) = r \]
can be rewritten as
\[ f(t) = b \fl r^* \]
where $b$ is a guard conveying all the matching information and $r^*$
is obtained from $b$ replacing every occurrence of a variable in the
left hand side by an application of a selector function.
In more detail, $f(t_1,\ldots,t_n) = r$ is rewritten as 
\[ f(t) = match(t_1,t) \wedge \cdots \wedge match(t_n,t) \fl
          \delta(r,t)
\]
where
\begin{eqnarray*}
match(K,t)                 &=& isK(t)\\
match(C(w_1,\ldots,w_m),t) &=& isC(t) \wedge
                               match(w_1,t|_1) \wedge \cdots \wedge
                               match(w_m,t|_m)          \\
match(x,t)                 &=& True
\end{eqnarray*}
and
\begin{eqnarray*}
\delta(f(e_1,\ldots,e_j),t)  &=& f(\delta(e_1,t),\ldots,\delta(e_j,t))\\
\delta(x,t)                  &=& \texttt{proj}(pos(x,t))(t)
\end{eqnarray*}
provided that $\texttt{proj}(p)(t)$ returns $t|_p$ and that $pos(x,t)$
is the position where $x$ occurs at $t$.  

\item Given a rule
      \[ f(t_1,\ldots,t_n) = b \fl r \]
      either $r$ is a variable, or a constant, or it 
      can be rewritten as
      \[ f(\bart) = b \fl g(h_1(\bart),\ldots,h_m(\bart)) \]
      Moreover, this is true for the whole set of rules in a program,
      i.e.~the newly introduced function definitions 
      -- for $h_1,\ldots,h_m$ -- can again be normalized and the whole
      transformation is terminating.
\end{enumerate}
\end{lemma}

\begin{myproof}[Main Function Constraint]
From lemma~\ref{lemma:transform}.1, the definition of $(\sqcup)$ and the
semantics of the language,
\[ p \circ f = p \circ f.1 \sqcup \cdots \sqcup p \circ f.n \]
for every partial predicate $p$ and function symbol $f$.
\end{myproof}

\begin{myproof}[Main Rule Constraint]
This is one of the lossy steps. This is justified by a generic
property of projections: if 
\[ p \in \PP~(\tau_1 \times \cdots \times \tau_n) \] 
then
\[ p \sqsubset p_1 \times \cdots \times p_n \]
so if $f.i$ is a function on tuples
\[ p\circ f.i \sqsubset 
                (p\circ f.i)_1 \times \cdots \times (p\circ f.i)_n \]
%\emph{(Aquí queda algún cabo por atar; tengo una solución basada en una
%  semántica no estándar, pero no acaba de gustarme.)}
\end{myproof}

%% \begin{example}
%% The rules for \emph{plus} can be rewritten as
%% %
%% %\begin{small}
%% \begin{verbatim}
%% plus1 = sel_m1          plus2 = Succ . plus . (cross (sel_n, sel_m2))

%% sel_m1 (Zero, m) = m    sel_n (Succ n, _) = n
%%                         sel_m2 (_, m) = m        cross (f,g) (x,y) = (f x, g y)%
%% \end{verbatim}
%% %\end{small}
%% \end{example}
%
\begin{myproof}[Head Constraints]
Given program rule 
\[ f.i~(t_1, \ldots, t_n) = r_i \]
we have to prove the inequation
\[ p\oo f.i.k \LT \Delta(p,f,i.k,t_k) \]
and, in fact, we are going to prove the equality.
Remember that $p\oo f.pos$ is intended to represent the projection at
position $pos$ of $p\circ f$. From lemma~\ref{lemma:transform}.2, 
we have
\[ f.i(t) = match(t_1,t) \wedge \cdots \wedge match(t_n,t) \fl
          \delta(r_i,t)\]
so 
\[ p\circ f.i(t) = match(t_1,t) \wedge \cdots \wedge match(t_n,t) \fl
          p(\delta(r_i,t)) \]
and from Def.~\ref{def:ppprojections}
\begin{eqnarray*}
\ppprj{k}{}(p\circ f.i)~x_k &=&
   p(f.i(x_1,\ldots,x_n)) \fl True \\
                            &=&  
   match(t_k,\barx) \wedge \cdots \wedge p(\delta(r_i,t))(\barx)
\end{eqnarray*}
where only $x_k$ and its subterms contribute to the result. 
From the definition of $\Delta$ and $\delta$ it can be proved that
\[ \Delta(p,f,i.k,t_k)~x_k = \ppprj{k}{}(p\circ f.i)~x_k \]
%% OJO!!
%% Ahora mismo no puedo dar nada mejor!!
%% Lo de que la desigualdad fuera realmente una igualdad me ha 
%% pillado completamente desprevenido!!
\end{myproof}

\begin{myproof}[Body Constraints]
In Sec.~\ref{reglas} three cases were considered:

\begin{enumerate}

\item \emph{(The body is a variable)}
The inequality to prove is
\[ p\oo f.i.k \LT p \]
provided that the rule for $f.i$ is of the form
\[ f.i(t_1,\ldots,t_n) = x \]
and that $x$ occurs at position $k$ at the head of the rule.

From lemma~\ref{lemma:transform}.2, we have that the definition of
$f.i$ can always be cast as:
\begin{eqnarray*}
f.i(t) &=& match(t_1,t) \wedge \cdots \wedge match(t_n,t) \fl
         \texttt{proj}(k)(t)
\end{eqnarray*}
and then
\begin{eqnarray*}
p\circ f.i(t) &=& match(t_1,t) \wedge \cdots \wedge match(t_n,t) \fl
         p(\texttt{proj}(k)(t))
\end{eqnarray*}
so
\begin{eqnarray*}
\ppprj{k}{}(p\circ f.i)~x_k &=& match(t_1,t) \wedge \cdots \wedge 
                                match(t_n,t) \wedge p(x_k)
\end{eqnarray*}
which is clearly less defined or equal than $p$, hence the inequality.

\item \emph{(The body is a constant)} The proof is very similar to
      that of the last case.

\item \emph{(The body is a function application)}
Without loss of generality (see lemma~\ref{lemma:transform}.3), 
we will restrict ourselves to rules of the form
\[ f(t) = b \fl g(h_1(t), \ldots, h_m(t)) \]
so
\[ (p\circ f)(t) = b \fl (p\circ g)(h_1(t), \ldots, h_m(t)) \]
To show the correctness of the inequalities
\begin{eqnarray*}
p\_ f.1 &\LT& (p\oo g)_1 \\
         &\vdots&         \\
p\_ f.m &\LT& (p\oo g)_m \\
p\oo f &\LT& p\_ f.1\oo h_1 \\
         &\vdots&               \\
p\oo f &\LT& p\_ f.m\oo h_m
\end{eqnarray*}
we can proceed by \emph{reductio ad absurdum}. As the $p\_f.i.j$
variables are only constrained by these inequalities, we can be sure
that they will take the greatest values. The only possibility for the
system to fail is that one inequality in the second set fails.
Let $z$ be an element such that $z \in (p\circ f)$ and 
$z \notin (p\_ f.k\oo h_k)\,$.
Introducing $z$ in the equation above for $(p\circ f)$ leads to
immediate contradiction. 
\end{enumerate}
\end{myproof}

\begin{myproof}[Simplification]
Trivial, as this is essentially replacement of equals by equals.
\end{myproof}

\begin{myproof}[Weakening]
Trivial as this is essentially replacement of a term by a greater one
in the right hand side of ``lesser than'' inequation.
\end{myproof}

%%% Local Variables: 
%%% mode: latex
%%% TeX-master: "tplp-reviewed"
%%% End: 

\end{document}